\documentclass[a4paper]{article}
\usepackage[utf8]{inputenc}
\usepackage[T1]{fontenc}
\usepackage{lmodern}
\usepackage[english]{babel}
\usepackage{a4wide}
\usepackage[affil-it]{authblk}
\usepackage[automark]{scrpage2}
\usepackage{graphicx}
\usepackage{verbatim}
\usepackage{caption}
\usepackage{subcaption}
\usepackage{sidecap}
\usepackage{blindtext}
\usepackage{cancel}
\usepackage{amsmath}
\usepackage{amssymb}
\usepackage{dsfont}
\usepackage{tensor}
\usepackage{slashed}
\usepackage{multirow}
\usepackage{float}
\usepackage[colorlinks,pdfpagelabels, pdfstartview = FitH, bookmarksopen =
true, bookmarksnumbered = true, linkcolor = blue, plainpages = false,
hypertexnames = false, citecolor = blue]{hyperref}
\usepackage[toc,page]{appendix}
\usepackage{cite}
\usepackage{amsfonts}
\numberwithin{equation}{section}
\allowdisplaybreaks[2]
\restylefloat{table}
\begin{document}

	\title{Phenomenology of pseudotensor mesons and the pseudotensor glueball}
	\author{Adrian Koenigstein\textsuperscript{1,2}, Francesco Giacosa\textsuperscript{3,1}}

	\affil{\textsuperscript{1}Institute for Theoretical Physics, Johann Wolfgang Goethe University,\\ Max-von-Laue-Str.\ 1, 60438 Frankfurt am Main, Germany}

	\affil{\textsuperscript{2}Frankfurt Institute for Advanced Studies,\\ Ruth-Moufang-Str.\ 1, 60438 Frankfurt am Main, Germany}

	\affil{\textsuperscript{3}Institute of Physics, Jan Kochanowski University,\\ ul. Swietokrzyska 15, 25-406 Kielce, Poland}

	\maketitle

\begin{abstract}

	We study the decays of the pseudotensor mesons $[ \pi_{2}(1670) ,\, K_{2}(1770) ,\, \eta_{2}(1645) ,\, \eta_{2}(1870) ]$ interpreted as the ground-state nonet of $1^{1}D_{2}$ $\bar{q}q$ states using interaction Lagrangians which couple them to pseudoscalar, vector, and tensor mesons. While the decays of $\pi_{2}(1670)$ and $K_{2}(1770)$ can be well described, the decays of the isoscalar states $\eta_{2}(1645)$ and $\eta_{2}(1870)$ can be brought in agreement with the present experimental data only if the mixing angle between nonstrange and strange states is surprisingly large (about $-42^{\circ}$, similar to the mixing in the pseudoscalar sector, in which the chiral anomaly is active). Such a large mixing angle is however at odd with all other conventional quark-antiquark nonets: if confirmed, a deeper study of its origin will be needed in the future. Moreover, the $\bar{q}q$ assignment of pseudotensor states predicts that the ratio $[ \eta_{2}(1870) \rightarrow a_{2}(1320)\,\pi]/[\eta_{2}(1870) \rightarrow f_{2}(1270)\,\eta]$ is about $23.5$. This value is in agreement with Barberis et al., ($20.4 \pm 6.6$), but disagrees with the recent reanalysis of Anisovich et al., ($1.7 \pm 0.4$). Future experimental studies are necessary to understand this puzzle. If Anisovich's value shall be confirmed, a simple nonet of pseudoscalar mesons cannot be able to describe data (different assignments and/or additional state, such as an hybrid state, will be needed). In the end, we also evaluate the decays of a pseudoscalar glueball into the aforementioned conventional $\bar{q}q$ states: a sizable decay into $K^{\ast}_{2}(1430)\,K$ and $a_{2}(1230)\,\pi$ together with a vanishing decay into pseudoscalar-vector pairs [such as $\rho(770)\,\pi$ and $K^{\ast}(892)\,K$] are expected. This information can be helpful in future studies of glueballs at the ongoing BESIII and at the future PANDA experiments.
	
\end{abstract}


\section{Introduction}
\label{sec:introduction}

	Most of the light mesons listed in the Particle Data Group (PDG) \cite{pdg} can be understood as quark-antiquark ($\bar{q}q$) states \cite{isgur}, see also e.g. the review in\ Ref. \cite{klemptrev}. Quark-antiquark mesons, also denoted as conventional mesons, are grouped into nonets. This is a consequence of flavor symmetry, $U_{V}(3)$: in the limit of equal masses of the light quark flavors $u$, $d$, and $s$, the QCD Lagrangian is invariant under their interchange. In reality, this symmetry is explicitly broken by non-equal bare quark masses, mostly due to the fact that the strange quark $s$ is sizable heavier than the up and down quarks $u$ and $d$ \cite{pdg}.\\
	In the limit of vanishing light quark masses, the QCD Lagrangian exhibits also chiral symmetry, $U_{R}(3)\times U_{L}(3)$. Quark-antiquark nonets with equal total angular momentum $J$ but with opposite parity $P$ are connected by chiral transformations. For instance, scalar ($1^3 P_0$ , $J^{PC}=0^{++}$) and pseudoscalar mesons ($1^1 S_0$ , $J^{PC}=0^{-+}$) as well as vector
	($1^3 S_1$ , $J^{PC}=1^{--}$) and axial-vector mesons ($1^3 P_1$ , $J^{PC}=1^{++}$) are chiral partners, e.g. Ref. \cite{dick}. In addition to explicit breaking, chiral symmetry is -- even more importantly -- also broken spontaneously, $U_{R}(3) \times U_{L}(3) \rightarrow U_{V}(3)$: pseudoscalar mesons (e.g., the pions) are the corresponding quasi-Goldstone bosons.\\
	Tensor mesons ($1^3 P_{2}$ , $J^{PC}=2^{++}$) are another example of a very well-understood $\bar{q}q$ nonet: their decays fit nicely into this scheme \cite{isgur,burakovsky,tensor,ciricgliano}. The chiral partners of tensor mesons, the pseudotensor mesons ($1^1 D_{2}$, $J^{PC}=2^{-+}$), are not so well understood, e.g. Refs. \cite{wang,bing} and Refs. therein. The standard assignment \cite{pdg,isgur} contains the isotriplet state $\pi_{2}(1670),$ the isodoublet states $K_{2}(1770),$ and the isoscalar states $\eta_2(1645)$ and $\eta_2(1870)$. We plan to study the decays of these resonances in order to test the validity of this assignment and to investigate the mixing in the isoscalar sector. To this end, we build two effective interaction Lagrangians which describe the decays of pseudotensor states into vector-pseudoscalar and into tensor-pseudoscalar pairs. The isotriplet and isodoublet states $\pi_{2}(1670)$ and $K_{2}(1770)$ fit well into the $\bar{q}q$ picture. However, in the isoscalar sector the situation is not that simple: the only possible way to understand the experimental results of $\eta_2(1645)$ and $\eta_2(1870)$ listed in the PDG is a very large mixing in the isoscalar-pseudotensor sector. In fact, the mixing angle between the purely nonstrange and strange states turns out to be about $-42^{\circ}$ in our model, which is very similar to the one in the pseudoscalar sector (leading to the famous mixing among $\eta$ and $\eta^{\prime}(958)$, e.g. Ref. \cite{AmelinoCamelia:2010,feldmann,thooftanomaly}.) However, while the large pseudoscalar-isoscalar mixing is related to the chiral anomaly, such an analogous strong mixing in the pseudotensor sector, if confirmed, would require a careful analysis in order to be understood.\\
	In addition to the mixing angle, in the isoscalar sector there are some conflicting experimental informations about the ratio $[\eta_{2}(1870) \rightarrow a_{2}(1320)\,\pi]/[\eta_{2}(1870) \rightarrow f_{2}(1270)\,\eta]$ that we will discuss. From an experimental point of view it is expected that the GlueX \cite{gluex1,gluex2,gluex3} and CLAS12 \cite{clas12} experiment at Jefferson Lab will soon start taking data (photoproduction of mesons by photon-nucleon scattering) and can possibly shed light on pseudotensor states in general, with special attention to the $\eta_2(1870)$ state. Indeed, one of the main motivations of the present work is to draw attention on the existing problems in understanding pseudotensor states. The hope is that new and better experimental data from Jlab will be  available in the upcoming years.\\
	Another interesting -- although yet only theoretical -- pseudotensor state is the pseudotensor glueball. According to lattice QCD its mass is supposed to be about $3$ GeV \cite{mainlattice,gregory}. A simple modification of our interaction Lagrangians allows to make some predictions concerning this putative state. This might help to identify possible glueball candidates.\\
	The article is organized as follows: in Sec.\ \ref{sec:two} we discuss $\bar{q}q$ nonets, their microscopic currents and quantum numbers; then, we introduce the two effective Lagrangians mentioned above and the corresponding decay widths. In\ Sec.\ \ref{sec:results} we present the results for the decays of isotriplet and isodoublet members of the pseudoscalar nonet. In\ Sec.\ \ref{sec:isoscalarmixing} we turn to the isoscalar sector: after showing that a small mixing angle cannot be correct, we show that a large negative mixing angle allows to understand all the existing data. Within this context, we describe also some puzzling entries in the PDG.\ In Sec.\ \ref{sec:glueball} we concentrate on the decays of a (yet hypothetical) pseudoscalar glueball. Finally, in\ Sec.\ \ref{sec:conclusions} we turn to the conclusions and outlooks.


\section{The model}
\label{sec:two}

	In this Section we describe the model for the decays of pseudotensor mesons. First, we introduce the relevant quark-antiquark nonets and provide a brief repetition of some aspects of the quark-antiquark assignments. In the second step, the effective Lagrangians for the present model are presented.

\subsection{Quark-antiquark nonets}

	The underlying structure of conventional meson fields is described via $\bar{q}q$ currents. First, we briefly review their quantum numbers and illustrate their construction scheme.\\
	The total spin of a $\bar{q}q$ bound state can be either $S=0$ or $S=1$, whereas the angular momentum is not limited: $L = 0\, \text{(ground state)},\, 1 ,\, 2 ,\, 3 ,\, \ldots\,$. The total angular momentum $\vec{J}=\vec{L}+\vec{S}$ is restricted by $\left\vert L-S\right\vert\leq J\leq L+S$. Parity and charge conjugation quantum numbers are given by
	
	\begin{align}
		P & =(-1)^{L+1}\,, && C = (-1)^{L+S} \,,
	\end{align}
		
	(where -- strictly speaking -- only diagonal members of a given multiplet are $C$-eigenstates). As usual, mesons are grouped into nonets (see below) corresponding to a well-defined combination $J^{PC}$ \cite{pdg}. In Tab.\ (\ref{tab:quarkcurrents}) we present various combinations $J^{PC}$ together with the values of $L,$ $S,$	the old spectroscopic notation $n$ $^{2S+1}L_{J},$ and the corresponding microscopic currents generating them (here, the quark fields reads $q_{i}$, where $i \in \{u,\,d,\,s\}$). We recall that $n$ is the radial quantum number ($n=1$ for all the cases under considerations), $L = S ,\, P ,\, D ,\,, \ldots\,$, and $\overleftrightarrow{\partial}_{\mu}=\overrightarrow{\partial}_{\mu}-\overleftarrow{\partial}_{\mu}$.
	
	\begin{table}[H] 
		\centering\renewcommand{\arraystretch}{2.0}
		\begin{tabular}[c]{|c|cc|c|c||c|}
			\hline
			Meson & $n^{2S+1}L_J$ & $J^{PC}$ & $S$ & $L$ & Hermitian quark current operators \\
			\hline
			pseudoscalar & $1^1 S_0$ & $0^{-+}$ & $0$ & \multirow{2}{*}{$0$} & $P_{ij} = \bar{q}_j \, i \gamma^{5} \, q_i$ \\
			\cline{1-4} \cline{6-6}
			vector & $1^3 S_1$ & $1^{--}$ & $1$ &  & $V_{ij}^{\mu} = \bar{q}_j \, \gamma^{\mu} \, q_i$ \\
			\hline \hline
			pseudovector & $1^1 P_1$ & $1^{+-}$ & $0$ & \multirow{4}{*}{$1$} & $P_{ij}^{\mu} = \bar{q}_j \, \gamma^{5} \overleftrightarrow{\partial}^{\mu} \, q_i$ \\
			\cline{1-4} \cline{6-6}
			scalar & $1^3 P_0$ & $0^{++}$ & $1$ &  & $S_{ij} = \bar{q}_j \, q_i$ \\
			\cline{1-4} \cline{6-6}
			axial vector & $1^3 P_1$ & $1^{++}$ & $1$ &  & $A_{ij}^{\mu} = \bar{q}_j \, \gamma^{5} \gamma^{\mu} \, q_i$ \\ 
			\cline{1-4} \cline{6-6}
			tensor & $1^3 P_2$ & $2^{++}$ & $1$ &  & $X_{ij}^{\mu\nu} = \bar{q}_j \, i \Big[ \gamma^{\mu} \overleftrightarrow{\partial}^{\nu} + \gamma^{\nu} \overleftrightarrow{\partial}^{\mu} - \frac{2}{3}\, \tilde{G}^{\mu\nu} \overleftrightarrow{\slashed{\partial}} \Big] \, q_i$ \\
			\hline \hline
			pseudotensor & $1^1 D_2$ & $2^{-+}$ & $0$ & \multirow{4}{*}{$2$} & $T_{ij}^{\mu\nu} = \bar{q}_j \, i \Big[ \gamma^{5} \overleftrightarrow{\partial}^{\mu} \overleftrightarrow{\partial}^{\nu} - \frac{2}{3}\, \tilde{G}^{\mu\nu} \overleftrightarrow{\partial}_{\alpha} \overleftrightarrow{\partial}^{\alpha} \Big] \, q_i$ \\
			\cline{1-4} \cline{6-6}
			excited vector & $1^3 D_1$ & $1^{--}$ & $1$ &  & $S_{ij}^{\mu} = \bar{q}_j \, \overleftrightarrow{\partial}^{\mu} \, q_i$ \\
			\cline{1-4} \cline{6-6}
			axial tensor & $1^3 D_2$ & $2^{--}$ & $1$ &  & $B_{ij}^{\mu\nu} = \bar{q}_j \, i \Big[ \gamma^{5} \gamma^{\mu} \overleftrightarrow{\partial}^{\nu} + \gamma^{5} \gamma^{\nu} \overleftrightarrow{\partial}^{\mu} - \frac{2}{3}\, \tilde{G}^{\mu\nu} \gamma^{5} \overleftrightarrow{\slashed{\partial}} \Big] \, q_i$ \\
			\cline{1-4} \cline{6-6}
			spin-3 tensor & $1^3 D_3$ & $3^{--}$ & $1$ &  & \ldots \\
			\hline
		\end{tabular}
		\caption{Types of different mesons and their corresponding quantum numbers. We use the projector $\tilde{G}^{\mu\nu} = \eta^{\mu\nu} - \frac{k^\mu k^\nu}{k^2}$.}
		\label{tab:quarkcurrents}
	\end{table}

	We now introduce the matrices of the nonets relevant in the present work. The nonet of pseudoscalar mesons is given by

	\begin{align}
		P & =
		\begin{pmatrix}
			\frac{\eta_{N} + \pi^{0}}{\sqrt{2}} & \pi^{+} & K^{+} \\
			\pi^{-} & \frac{\eta_{N} - \pi^{0}}{\sqrt{2}} & K^{0} \\
			K^{-} & \bar{K}^{0} & \eta_{S}
		\end{pmatrix} \,, \label{eq:pseudoscalar_nonet}
	\end{align}
	
	where $\eta_{N}=\sqrt{1/2}\,(\bar{u}u+\bar{d}d)$ is the pure non-strange state and $\eta_{S}=\bar{s}s$ the pure strange state. The identifications of the fields with physical resonances are listed in Tab.\ (\ref{tab:resonances_states}), while the	transformation properties under parity, charge conjugation, and flavor transformations are reported in Tab.\ (\ref{tab:transformations}). The isoscalar mixing angle $\beta_{p} = -40.5^{\circ}$ in the non-strange-strange pseudoscalar sector is taken from \cite{AmelinoCamelia:2010}. The large maxing is linked to the chiral anomaly, see e.g. Refs. \cite{klemptrev,feldmann,dick}.
	
	\begin{table}[H] 
		\centering
		\renewcommand{\arraystretch}{1.1}
		\begin{tabular}[c]{|l|l|}
			\hline
			Physical resonances & Nonet $q\bar{q}$-states \\
			\hline \hline
			$\pi$ & $\hphantom{-}\pi$ \\
			$K$ & $\hphantom{-}K$ \\
			$\eta$ & $\hphantom{-}\eta_{N} \cos\beta_{p} + \eta_{S} \sin\beta_{p}$ \\
			$\eta^{\prime}(958)$ & $- \eta_{N} \sin\beta_{p} + \eta_{S} \cos\beta_{p}$ \\
			\hline
			$\rho(770)$ & $\hphantom{-}\rho$ \\
			$K^{*}(892)$ & $\hphantom{-}K^{*}$ \\
			$\omega(782)$ & $\hphantom{-}\omega_{N} \cos\beta_{v} + \omega_{S} \sin\beta_{v}$ \\
			$\phi(1020)$ & $-\omega_{N} \sin\beta_{v} + \omega_{S} \cos\beta_{v}$ \\
			\hline
			$a_{2}(1320)$ & $\hphantom{-}a_{2}$ \\
			$K_{2}^{*}(1430)$ & $\hphantom{-}K_{2}^{*}$ \\
			$f_{2}(1270)$ & $\hphantom{-} f_{2,N} \cos\beta_{t} + f_{2,S} \sin\beta_{t}$ \\
			$f^{\prime}_{2}(1525)$ & $- f_{2,N} \sin\beta_{t} + f_{2,S} \cos\beta_{t}$ \\
			\hline
			$\pi_{2}(1670)$ & $\hphantom{-} \pi_{2}$ \\
			$K_{2}(1770)$ & $\hphantom{-}K_{2}$ \\
			$\eta_{2}(1645)$ & $\hphantom{-} \eta_{2,N} \cos\beta_{pt} + \eta_{2,S} \sin\beta_{pt}$ \\
			$\eta_{2}(1870)$ & $- \eta_{2,N} \sin\beta_{pt} + \eta_{2,S} \cos\beta_{pt}$ \\
			\hline
		\end{tabular}
		\caption{Assignment of physical resonances to $\bar{q}q$-states in the model.}
		\label{tab:resonances_states}
	\end{table}
		
	Similarly, vector mesons are arranged in the vector nonet given by
	
	\begin{align}
		V^{\mu} & =
		\begin{pmatrix}
			\frac{\omega_{N}^{\mu} + \rho^{0\mu}}{\sqrt{2}} & \rho^{+\mu} & K^{\ast+\mu} \\
			\rho^{-\mu} & \frac{\omega_{N}^{\mu} - \rho^{0\mu}}{\sqrt{2}} & K^{\ast0\mu} \\
			K^{\ast-\mu} & \bar{K}^{\ast0\mu} & \omega_{S}^{\mu}
		\end{pmatrix} \,, \label{eq:vector_nonet}
	\end{align}
	
	where $\omega_{N}$ is purely non-strange and $\omega_{S}$ purely strange. Identifications and transformation properties are provided by Tab.\ (\ref{tab:resonances_states}) and Tab.\ (\ref{tab:transformations}). The isoscalar-vector mixing angle is very small: $\beta_{v}=-3.8^{\circ}$ \cite{pdg}. The physical states $\omega$ and $\phi$ are dominated by non-strange and strange components, respectively.\\	
	We now turn to tensor and pseudotensor states (for a recent review on mathematical aspects of tensor fields, see Ref. \cite{Koenigstein:2015} and references therein). The tensor meson nonet reads
	
	\begin{align}
		X^{\mu\nu} & =
		\begin{pmatrix}
			\frac{f_{2,N}^{\mu\nu} + a_{2}^{0\mu\nu}}{\sqrt{2}} & a_{2}^{+\mu\nu} &	K_{2}^{\ast+\mu\nu} \\
			a_{2}^{-\mu\nu} & \frac{f_{2,N}^{\mu\nu} - a_{2}^{0\mu\nu}}{\sqrt{2}} &	K_{2}^{\ast0\mu\nu} \\
			K_{2}^{\ast-\mu\nu} & \bar{K}_{2}^{\ast0\mu\nu} & f_{2,S}^{\mu\nu}
		\end{pmatrix} \,. \label{eq:tensor_nonet}
	\end{align}
	
	In analogy to the vector case, the mixing angle between the pure non-strange $f_{2,N}$ and the pure strange $f_{2,S}$ is small: $\beta_{t}=3.2^{\circ}$, see Ref. \cite{pdg}. Finally, the nonet of pseudotensor states -- which constitutes the main subject of the present work -- is given by
		
	\begin{align}
		T^{\mu\nu} & =
		\begin{pmatrix}
			\frac{\eta_{2,N}^{\mu\nu} + \pi_{2}^{0\mu\nu}}{\sqrt{2}} & \pi_{2}^{+\mu\nu} & K_{2}^{+\mu\nu} \\
			\pi_{2}^{-\mu\nu} & \frac{\eta_{2,N}^{\mu\nu} - \pi_{2}^{0\mu\nu}}{\sqrt{2}} & K_{2}^{0\mu\nu} \\
			K_{2}^{-\mu\nu} & \bar{K}_{2}^{0\mu\nu} & \eta_{2,S}^{\mu\nu}
		\end{pmatrix} \,, \label{eq:pseudotensor_nonet}
	\end{align}
	
	see again Tab.\ (\ref{tab:resonances_states}) and Tab.\ (\ref{tab:transformations}) for the physical content and transformations of (pseudo)tensor mesons.\\
	The mixing angle $\beta_{pt}$ is -- at first -- unknown. Naively, one would	expect a small mixing angle (similarly to the vector and the tensor mesons;	alternatively, one may use the Okubo formula, which yields $\beta_{pt} \approx 14.8^{\circ}$ derived from (15.9) in \cite{pdg}). However, a small mixing angle does not lead to acceptable results within our model. A large and negative mixing angle -- as in the pseudoscalar sector -- is needed, see Sec.\ \ref{sec:isoscalarmixing}.\\
	\\
	Other mesonic nonets can be constructed in the same way, see for instance Refs. \cite{dick,florianlisa,julia}. They are omitted here, since they do not enter the decays under consideration.

	\begin{table}[H] 
		\centering
		\renewcommand{\arraystretch}{1.1}
		\begin{tabular}[c]{|c|c|c|c|}
			\hline
			Nonet & Parity $(P)$ & Charge congugation $(C)$ & Flavour $(U_{V}(3))$ \\
			\hline \hline
			$P$ & $-P(t,-\vec{x})$ & $P^{t}$ & $U P U^{\dagger}$ \\
			\hline
			$V^{\mu}$ & $V_{\mu}(t,-\vec{x})$ & $-(V^{\mu})^{t}$ & $U V^{\mu}U^{\dagger}$ \\
			\hline
			$X^{\mu\nu}$ & $X_{\mu\nu}(t,-\vec{x})$ & $(X^{\mu\nu})^{t}$ & $U X^{\mu\nu} U^{\dagger}$ \\
			\hline
			$T^{\mu\nu}$ & $-T_{\mu\nu}(t,-\vec{x})$ & $(T^{\mu\nu})^{t}$ & $U T^{\mu\nu} U^{\dagger}$ \\
			\hline
		\end{tabular}
		\caption{Transformation properties of the pseudoscalar (\ref{eq:pseudoscalar_nonet}), the vector (\ref{eq:vector_nonet}), the tensor (\ref{eq:tensor_nonet}), and pseudotensor (\ref{eq:pseudotensor_nonet}) nonets under charge conjugation $C$, parity $P,$ and flavour transformations $U_{V}(3)$. Notice the position of the Lorentz indices for parity transformations, since spatial and time-like indices do not transform identically.}
		\label{tab:transformations}
	\end{table}


\subsection{The interaction Lagrangians and tree-level decay width}
\label{sec:themodel}

	Using the nonets (\ref{eq:pseudoscalar_nonet}) - (\ref{eq:pseudotensor_nonet}) introduced in the previous subsection, we construct two effective interaction Lagrangians which describe the decays of pseudotensor mesons.\\
	The first Lagrangian $\mathcal{L}_{TVP}$ includes the coupling of pseudotensor mesons to vector-pseudoscalar pairs, 

	\begin{align}
		\mathcal{L}_{TVP} & = c_{TVP}\, \mathrm{Tr} \big\{  T_{\mu\nu} \big[  V^{\mu} ,\, (\partial^{\nu}P) \big]  _{-} \big\} \,, \label{eq:lagtvp}
	\end{align}
	
	where $c_{TVP}$ denotes an (at first) unknown dimensionless coupling constant and $[\,,]_{-}$ is the commutator.\\
	The second interaction Lagrangian $\mathcal{L}_{TXP}$ contains the coupling of pseudotensor mesons to tensor-pseudoscalar pairs,
	
	\begin{align}
		\mathcal{L}_{TXP} & = c_{TXP}\, \mathrm{Tr} \big(  T_{\mu\nu} \big\{  X^{\mu\nu} ,\, P \big\}_{+} \big) \,, \label{eq:lagtxp}
	\end{align}
	
	where $c_{TXP}$ is an (at first) unknown coupling constant with dimension energy and $\{\,,\}_{+}$ is the anti-commutator.\\
	Both Lagrangians are invariant under $CPT$- , Poincar\'{e}- and flavour-transformations listed in Tab.\ (\ref{tab:transformations}). The explicit form of the Lagrangians are presented in App.\ \ref{app:lagrangians}. Hence, the total Lagrangian, which specifies our model, is given by
		
	\begin{align}
		\mathcal{L}_{T-\text{total}} & = \mathcal{L}_{\text{kin}} + \mathcal{L}_{TVP} + \mathcal{L}_{TXP} \,, \label{eq:ltot}
	\end{align}

	where $\mathcal{L}_{\text{kin}}=\frac{1}{2} \mathrm{Tr}\big[(\partial_{\mu}P)(\partial^{\mu}P)\big] + \ldots$ contains the usual kinetic terms of all relevant nonets (for the tensor fields, see again Ref. \cite{Koenigstein:2015}), while the interaction term is the sum of the two interaction Lagrangians presented above.\\
	The interaction Lagrangians outlined above can be considered as the dominant terms in the large-$N_{c}$ and flavour breaking expansions. For instance, the full Lagrangian for the interaction of the pseudotensor field with tensor and pseudoscalar field takes the form
	
	\begin{align}
		\mathcal{L}_{TXP}^{full} & = c_{TXP}\, \mathrm{Tr} \big(T_{\mu\nu} \big\{ X^{\mu\nu} ,\,P \big \}_{+} \big) + c_{TXP}\, \mathrm{Tr} \big( \hat{\delta}\, T_{\mu\nu} \big\{ X^{\mu\nu},\,P \big\}_{+} \big) + \label{ltxpfull} \vphantom{\frac{1}{2}}
		\\
		& \quad\, + \tilde{c}_{TXP}\, \mathrm{Tr} \big( T_{\mu\nu} \big)\, \mathrm{Tr} \big( \big\{ X^{\mu\nu},\,P\big \}_{+} \big) + \tilde{c}_{TXP}\, \mathrm{Tr} \big( \hat{\delta}\, T_{\mu\nu} \big)\, \mathrm{Tr} \big( \big\{X^{\mu\nu},\,P\big\}_{+} \big) + \ldots \vphantom{\frac{1}{2}} \nonumber
	\end{align}
	
	where the matrix $\hat{\delta} = \mathrm{diag} ( 0,\, \delta_{d},\,\delta_{s} )$ describes isospin violation. Following \cite{Amsler:1995td} and denoting $V$ as the potential responsible of the creation of a quark-antiquark pair from the QCD vacuum, we find,
	
	\begin{align*}
		& \delta_{d} =\frac{ \langle 0 | V | \bar{d} d \rangle }{\langle 0 | V | \bar{u} u \rangle } - 1 \simeq 0 &&\text{(isospin violation can be neglected to a very good level of accuracy),}
		\\
		& \delta_{s} = \frac{ \langle 0 | V | \bar{s}s \rangle }{\langle 0 | V | \bar{u}u \rangle } - 1 \lesssim 0.2 && \text{(small but eventually important for a precise description).}
	\end{align*}

	A similar value for $\delta_{s}$ has been obtained in Refs.\ \cite{Giacosa:2004ug,isgur}. Moreover, according to Refs.\ \cite{largen1,largen2,largen3} we expect

		\begin{align}
			c_{TXP} & \propto N_{c}^{-1/2} \,,
			\\
			\tilde{c}_{TXP} & \propto N_{c}^{-3/2} \,.		
		\end{align}

	The first term in Eq.\ (\ref{ltxpfull}) is the dominant one: due to the present experimental status this is the only term in the pseudeotensor-tensor-pseudoscalar sector which is used in the numerical calculations of the present work. Then, the second and the third term of Eq.\ (\ref{ltxpfull}) generate decay amplitudes proportional to $c_{TXP}\,\delta_{s}$ and $\tilde{c}_{TXP}$ respectively, hence they generate corrections of the order of 5-10\%, which should be taken into account when better experimental data will be available.\ Further terms such as the fourth term generate a contribution of the type $\tilde{c}_{TXP}\, \delta_{s},$ hence they are suppressed twice, because of large-$N_{c}$ and flavour symmetry violation. Their influence is expected to be about $1\%$, hence negligible.\\
	For what concerns the interaction term of pseudotensor mesons with pseudoscalar and vector fields the expansion is similar:
	
	\begin{align}
		\mathcal{L}_{TVP}^{full} & = c_{TVP}\, \mathrm{Tr} \big\{ T_{\mu\nu} \big[V^{\mu},\,(\partial^{\nu}P)\big]_{-} \big\} + c_{TVP}\, \mathrm{Tr} \big\{\hat{\delta}\, T_{\mu\nu} \big[V^{\mu},\,(\partial^{\nu}P)\big]_{-} \big\} \vphantom{\frac{1}{2}} \label{ltvpfull}
		\\
		& \quad\, + \tilde{c}_{TVP}\, \mathrm{Tr} \big\{ \hat{\delta}\, T_{\mu\nu}\big\}\, \mathrm{Tr} \big\{ \hat{\delta}\, \big[V^{\mu},\,(\partial^{\nu}P)\big]_{-} \big\} + \ldots \vphantom{\frac{1}{2}} \nonumber
	\end{align}
	
	As before, the first term dominates and the only term of Eq.\ (\ref{ltvpfull}) considered in this paper. There is however a difference between Eq.\ (\ref{ltxpfull}) and Eq.\ (\ref{ltvpfull}): due to the anti-commutator in the latter, some terms do not appear in the expansion. The next-to-leading contribution is the second term $\propto c_{TVP}\, \delta_{s}$ (breaking of flavour symmetry). Furthermore, one has terms $\propto \tilde{c}_{TVP}\,\delta_{s}$ such as the third one which suppressed in $N_{c}$ and the flavour expansions.\\
	\\		
	We now turn to the tree-level decay widths. The tree-level decay widths derived from the two Lagrangians $\mathcal{L}_{TVP}$ and $\mathcal{L}_{TXP}$ read
	
	\begin{align}
		\Gamma_{T\rightarrow VP}^{tl}(m_{T},\,m_{V},\,m_{P}) & = \frac{k_f}{8\pi\,m_{T}} \frac{g_{TVP}^{2}}{15} \Big( 2\, \frac{k_f^{4}}{m_{V}^{2}} + 5\, k_f^2 \Big)\, \Theta( m_{T}-m_{V}-m_{P} ) \,, \label{eq:tvp}
	\end{align}
	
	and
	
	\begin{align}
		\Gamma_{T\rightarrow XP}^{tl}(m_{T},\,m_{X},\,m_{P}) & = \frac{k_f}{8\pi\,m_{T}} \frac{g_{TXP}^{2}}{45} \Big( 4\, \frac{k_f^{4}}{m_{X}^{4}} + 30\,\frac{k_f^{2}}{m_{X}^{2}} + 45 \Big)\, \Theta( m_{T}-m_{X}-m_{P} ) \,. \label{eq:txp}
	\end{align}
	
	Above, $m_{T}$ is the mass of a (decaying) pseudotensor fields, while $m_{V}$, $m_{P},$ and $m_{X}$ are the masses of the vector, pseudoscalar, and tensor states (the decay products). $\Theta(x)$ denotes the Heaviside step-function. The quantities $g_{TVP}$ and $g_{TXP}$, which are proportional to the coupling constants $c_{TVP}$ and $c_{TXP}$, have to be calculated for each decay channel separately by using the expressions in App.\ \ref{app:lagrangians}. Finally, the function $k_f \equiv k_{f}(m_{T},\,m_{V},\,m_{P})$ is the modulus of the momentum $\vec{k}_f$ of one of the outgoing particles. Its analytic expression in Eq.\ (\ref{eq:tvp}) reads:
	
	\begin{align}
		k_{f} ( m_{T} ,\, m_{V} ,\, m_{P} ) & = \frac{1}{2\,m_{T}} \sqrt{ m_{T}^{4} + ( m_{V}^{2} - m_{P}^{2} )^{2} - 2\, m_{T}^{2}\, ( m_{V}^{2} + m_{P}^{2} ) } \,, \label{eq:kf}
	\end{align}
	
	while its expression in Eq.\ (\ref{eq:txp}) is obtained by substituting $m_{V}\rightarrow m_{X}.$\\
	These decay widths (\ref{eq:tvp}) and (\ref{eq:txp}) are derived via Feynman rules under the use of the polarization vectors (tensors) and their corresponding completeness relations. For a derivation of the unpolarized invariant decay amplitudes see App.\ \ref{app:deacyamplitudes}. \newline Calculations are performed at tree-level, that is NLO effects due to quantum loops are not considered here. Loops are expected to be small when the ratio $\Gamma/(M-M_{th})$ is sufficiently small ($\leq0.2$,  where $\Gamma$ and $M$ are the decay width and the mass of the decaying state and $M_{th}$ is the lowest threshold \cite{lupo}). This condition is fulfilled for the pseudotensor mesons under study.


\section{Results for $\pi_{2}(1670)$ and $K_{2}(1770)$}
\label{sec:results}

	As a first step, we determine the coupling constants $c_{TVP}$ and $c_{TXP}$ using two -- well known -- experimental decay widths,
	
	\begin{align}
		\pi_{2}(1670) & \rightarrow \rho(770)\,\pi \,, && \text{and} && \pi_{2}(1670) \rightarrow f_{2}(1270)\,\pi \,.
	\end{align}
	
	Taking into account that for these two decays one has $g_{TVP}^{2} = 2\,(\sqrt{2}\,c_{TVP})^{2}$ and $g_{TXP}^{2}=2\,\cos\beta_{t}\,c_{TXP}^{2}$, the use of Eqs.\ (\ref{eq:tvp}), (\ref{eq:txp}), and (\ref{eq:tvplagexp}) implies that [see Tab.\ (\ref{tab:ergebnisohneisoscalar})],
	
	\begin{align}
		c_{TVP}^{2} & = 11.9 \pm 1.6 \,, && c_{TXP}^{2} = (15.1 \pm 1.0) \cdot 10^{6} \, \, \text{MeV}^{2} \,. \label{eq:cs}
	\end{align}
	
	The quoted errors are solely determined by the experimental errors on the decay widths (mass errors are small and negligible for our accuracy).\\
	Once the coupling constants are known, all the isotriplet and isodoublet decays of pseudotensor states are uniquely fixed. The results are shown in Tab.\ (\ref{tab:ergebnisohneisoscalar}). 

	\begin{table}[H] 
		\centering
		\renewcommand{\arraystretch}{1.1}
		\begin{tabular}[c]{|l|c|c|}
			\hline
			Decay process & Theory (MeV) & Experiment (MeV) \\
			\hline \hline
			$\pi_{2}(1670) \rightarrow\rho(770)\,\pi$ & $80.6 \pm 10.8$ & $80.6 \pm 10.8$ \\
			\hline
			$\pi_{2}(1670) \rightarrow f_{2}(1270) \,\pi$ & $146.4 \pm 9.7$ & $146.4 \pm 9.7$ \\
			\hline \hline
			$\pi_{2}(1670) \rightarrow\bar{K}^{*}(892)\,K + c.c. $ & $11.7 \pm 1.6$ & $10.9 \pm 3.7$ \\
			\hline
			$\pi_{2}(1670) \rightarrow\bar{K}_{2}^{*}(1430)\,K + c.c. $ & $0$ & \\
			\hline
			$\pi_{2}(1670) \rightarrow f^{\prime}_{2}(1525) \,\pi$ & $0.1 \pm0.1$ & \\
			\hline
			$\pi_{2}(1670) \rightarrow a_{2}(1320) \,\pi$ & $0$ & not seen \\
			\hline
			$\pi_{2}(1670) \rightarrow a_{2}(1320) \,\eta$ & $0$ & \\
			\hline
			$\pi_{2}(1670) \rightarrow a_{2}(1320) \,\eta^{\prime}(958) $ & $0$ & \\
			\hline \hline
			$K_{2}(1770) \rightarrow\rho(770)\,K $ & $22.2 \pm3.0$ & \\
			\hline
			$K_{2}(1770) \rightarrow\bar{K}^{*}(892)\,\pi$ & $25.5 \pm3.4$ & seen \\
			\hline
			$K_{2}(1770) \rightarrow\bar{K}^{*}(892)\,\eta$ & $10.5 \pm1.4$ & \\
			\hline
			$K_{2}(1770) \rightarrow\bar{K}^{*}(892)\,\eta^{\prime}(958) $ & $0$ & \\
			\hline
			$K_{2}(1770) \rightarrow\omega(782)\,K$ & $8.3 \pm1.1$ & seen \\
			\hline
			$K_{2}(1770) \rightarrow\phi(1020)\,K$ & $4.2 \pm0.6$ & seen \\
			\hline
			$K_{2}(1770) \rightarrow a_{2}(1320)\,K$ & $0$ & \\
			\hline
			$K_{2}(1770) \rightarrow\bar{K}_{2}^{*}(1430)\,\pi$ & $84.5 \pm5.6$ & dominant \\
			\hline
			$K_{2}(1770) \rightarrow\bar{K}_{2}^{*}(1430)\,\eta$ & $0$ & \\
			\hline
			$K_{2}(1770) \rightarrow\bar{K}_{2}^{*}(1430)\,\eta^{\prime}(958)$ & $0$ & \\
			\hline
			$K_{2}(1770) \rightarrow f_{2}(1270)\,K$ & $5.8 \pm0.4$ & seen \\
			\hline
			$K_{2}(1770) \rightarrow f^{\prime}_{2}(1525)\,K$ & $0$ & \\
			\hline
		\end{tabular}
		\caption{Decays of $I=1$ and $I=1/2$ pseudotensor states. The first two
		entries were used to determine the coupling constants of the model, see Eq.\ (\ref{eq:cs}). The total decay widths are
		$\Gamma^{\text{tot}}_{\pi_{2}(1670)} = (260 \pm9)\,\,\text{MeV}$ and $\Gamma^{\text{tot}}_{K_{2}(1770)} = (186 \pm 14) \,\, \text{MeV}$.}
		\label{tab:ergebnisohneisoscalar}
	\end{table}

	Following comments are in order:

	\begin{enumerate}
		\item We recall that the first two entries of Tab.\ (\ref{tab:ergebnisohneisoscalar}) were used to calculate the coupling constants of the model, see Eq.\ (\ref{eq:cs}). These values are extracted from the PDG using the branching ratio quoted by the PDG: $56.3 \pm 3.2 \%$ for $\pi_{2}(1670) \rightarrow f_{2}(1270)\,\pi$ and $31 \pm 4 \%$ for $\pi_{2}(1670) \rightarrow \rho\,\pi.$ However, a closer inspection of the performed experiments on the resonance $\pi_{2}(1670)$ shows that new experiments are needed to improve experimental knowledge. For instance, the ratio $\Gamma(\rho\,\pi) / [0.565\, \Gamma(f_{2}(1270)\,\pi)]$, whose PDG quoted average is $0.97\pm0.09,$ was determined only by two distinct results: $0.76\pm0.07\pm0.10$ by \cite{Chung:2002pu} and $1.01\pm0.05$ by \cite{Barberis:1998in}. While consistent among	each other, a new experimental determination would be very welcome for our theoretical approach, since the determination of the coupling constant follows from these experimental values.
		
		\item The prediction for the decay channel $\pi_{2}(1670) \rightarrow K^{\ast}(892)\,K$ is in agreement with the present value quoted by the PDG. However, it should be stressed that this value is extracted from a single experiment \cite{Armstrong:1981kc}: a new experimental determination of this quantity is then compelling for a better comparison.
		
		\item In our model $\pi_{2}(1670)$ does not couple to the $a_{2}(1320)\,\pi$ channel, see Eq.\ (\ref{eq:txplagexp}). Experimentally this decay could also be	excluded to a good accuracy (for more experimental details see Ref.\ \cite{kuhn}).
		
		\item The experimental total decay width of $\pi_{2}(1670)$ reads $\Gamma_{\pi_{2}(1670)}^{\text{exp,tot}} = (260 \pm 9) \,\,$MeV. The value in our model is $\approx 238.8\,\,$MeV. The reason why the sum of the theoretical and experimental decay modes in our model is slightly smaller than the experimental total width is due to the fact that the model does not include the decay $\pi_{2}(1670)\rightarrow f_{0}(500)\,\pi,$ which is $\approx 26\,\,$MeV. (Other channels, which are not present in our model, would also contribute. Nevertheless they are negligibly small \cite{pdg}). Taking this into account, the model is consistent with the total width.
		
		\item Experimentally, the decay channel $K_{2}(1770)\rightarrow\bar{K}_{2}^{\ast}(1430)\,\pi$ is dominant. The total decay width of $K_{2}(1770)$ is	$\Gamma_{K_{2}(1770)}^{\text{exp,tot}}=(186\pm14)\,\,\text{MeV}$. Theoretically, the $K_{2}(1770)\rightarrow\bar{K}_{2}^{\ast}(1430)\,\pi$ decay mode is the dominant one ($84.5\pm5.6$ MeV).
		
		\item The full theoretical decay width of $K_{2}(1770)$ amounts to $(162.0 \pm 15.4)\,\,$MeV which is compatible with the experimental value.
				
		\item Various branching ratios of $K_{2}(1770)$ can be calculated from Tab.\ (\ref{tab:ergebnisohneisoscalar}). At present, experimental results are missing (no average or fit is quoted by PDG \cite{pdg}). In this respect, our approach makes predictions for new future experimental measurements. In this context, it will also be possible to determine the mixing angle of the bare $1^{1}D_{2}$ and $1^{3}D_{2}$ configurations into the physical $K_{2}(1770)$ and $K_{2}(1820)$ states [so far, $K_{2}(1770)$ is dominated by $1^{1}D_{2}$].
		
		\item Most of the decays which are predicted by our model to vanish were consistently not seen in experiments. Yet, the decays $\pi_{2}(1670) \rightarrow a_{2}(1230)\,\eta$ and $K_{2}(1270) \rightarrow K^{\ast}(892)\,\eta$ are expected to be small but not zero. They were not yet measured, hence they represent a test of our approach as soon as new experimental data will be available.
		
	\end{enumerate}


\section{The Isoscalar sector: $\eta_{2}(1645)$ and $\eta_{2}(1870)$}
\label{sec:isoscalarmixing}

	In this section we present the results of the isoscalar sector of the pseudotensor mesons. In Subsec.\ \ref{subsec:small_strange_nonstrange_mixing} we show that a small mixing of $\eta_{2,N}$ and $\eta_{2,S}$ is not capable to describe the data. Thus, in Subec.\ \ref{subsec:large_strange_nonstrange_mixing} we allow for a large mixing angle: a value of about $-40^{\circ}$ turns out to be favored. In Subsec.\ \ref{subsec:branching} we discuss the experimental status of branching ratios with particular attention to $[\eta_{2}(1870) \rightarrow a_{2}(1320)\,\pi]/[\eta_{2}(1870) \rightarrow f_{2}(1270)\,\eta]$, for which conflicting experimental results exist.


\subsection{Small strange-nonstrange mixing angle}
\label{subsec:small_strange_nonstrange_mixing}

	First, we present the results of the decay widths of the isoscalar $I=0$ pseudotensor states in Tab.\ (\ref{tab:ergebnisisoscalar}) for a vanishing mixing angle ($\beta_{pt}=0^{\circ}$) and for a small but non-vanishing angle ($\beta_{pt}=14.8^{\circ},$ obtained via the Okubo formula \cite{pdg}). Both cases lead to inconsistent results. In fact, the decay $\eta_{2}(1645)\rightarrow a_{2}(1320)\,\pi$ turns out to be larger than $300\,\,$MeV, which is not acceptable, since it is sizable larger than the total experimental width $\Gamma_{\eta_{2}(1645)}^{\text{exp,tot}}=(181\pm11)\,\,$MeV. These results hold for any small mixing angle ($\left\vert \beta_{pt}\right\vert \lesssim30^{\circ}$).
	
	\begin{table}[H] 
		\centering
		\renewcommand{\arraystretch}{1.1}
		\begin{tabular}[c]{|l|c|c|c|}
			\hline
			Decay process & Theory (MeV) & Theory (MeV) & Experiment (MeV) \\
			& $(\beta_{pt} = 14.8^{\circ})$ & $(\beta_{pt} = 0.0^{\circ})$ & \\
			\hline \hline
			$\eta_{2}(1645) \rightarrow\bar{K}^{*}(892)\,K + c.c. $ & $3.2 \pm0.4$ & $8.6 \pm 1.1$ & seen \\
			\hline
			$\eta_{2}(1645) \rightarrow a_{2}(1320)\,\pi$ & \textcolor{red}{$315.6 \pm 21.2$} & \textcolor{red}{$337.8 \pm 22.6$} & \\
			\hline
			$\eta_{2}(1645) \rightarrow\bar{K}_{2}^{*}(1430)\,K + c.c. $ & $0$ & $0$ & \\
			\hline
			$\eta_{2}(1645) \rightarrow f_{2}(1270)\,\eta$ & $0$ & $0$ & not seen \\
			\hline
			$\eta_{2}(1645) \rightarrow f_{2}(1270)\,\eta^{\prime}(958) $ & $0$ & $0$ & \\
			\hline
			$\eta_{2}(1645) \rightarrow f^{\prime}_{2}(1525)\,\eta$ & $0$ & $0$ & \\
			\hline
			$\eta_{2}(1645) \rightarrow f^{\prime}_{2}(1525)\,\eta^{\prime}(958) $ & $0$ & $0$ & \\
			\hline \hline
			$\eta_{2}(1870) \rightarrow\bar{K}^{*}(892)\,K + c.c. $ & $60.1 \pm8.0$ & $45.6 \pm6.1$ & \\
			\hline
			$\eta_{2}(1870) \rightarrow a_{2}(1320)\,\pi$ & \textcolor{red}{$32.2 \pm 2.1$} & \textcolor{red}{$0$} & \\
			\hline
			$\eta_{2}(1870) \rightarrow\bar{K}_{2}^{*}(1430)\,K + c.c. $ & $0$ & $0$ & \\
			\hline
			$\eta_{2}(1870) \rightarrow f_{2}(1270)\,\eta$ & $2.5 \pm0.2$ & $0.1 \pm0.1$ & \\
			\hline
			$\eta_{2}(1870) \rightarrow f_{2}(1270)\,\eta^{\prime}(958) $ & $0$ & $0$ & \\
			\hline
			$\eta_{2}(1870) \rightarrow f^{\prime}_{2}(1525)\,\eta$ & $0$ & $0$ & \\
			\hline
			$\eta_{2}(1870) \rightarrow f^{\prime}_{2}(1525)\,\eta^{\prime}(958) $ & $0$ & $0$ & \\
			\hline
		\end{tabular}
		\caption{Decays of $I=0$ pseudo-tensor states. The total decay widths are $\Gamma_{\eta_{2}(1645)}^{\text{tot}}=(181\pm11)\,\,\text{MeV}$ and $\Gamma_{\eta_{2}(1870)}^{\text{tot}}=(225\pm14)\,\,\text{MeV}$.}
		\label{tab:ergebnisisoscalar}
	\end{table}
	
	In addition, the following comments are in order:

	\begin{enumerate}
		\item The PDG reports the following ratios for $\eta_{2}(1645)$:
		
		\begin{align}
			\Gamma^{\text{exp}} (\bar{K}\,K\,\pi)/\Gamma^{\text{exp}}(a_{2}(1320)\,\pi) & = 0.07 \pm 0.03 \,, \vphantom{\frac{1}{2}}
			\\
			\Gamma^{\text{exp}} (a_{2}(1320)\,\pi)/\Gamma^{\text{exp}}(a_{0}(980)\,\pi) & = 13.1 \pm 2.3 \,. \vphantom{\frac{1}{2}}
		\end{align}
		
		Thus, the channel $\eta_{2}(1645)\rightarrow a_{2}(1320)\,\pi$ is indeed expected to be dominant. This fact is well captured by the theoretical results of Tab.\ (\ref{tab:ergebnisisoscalar}), which however overshoot the data. Moreover, the experiment shows that $\Gamma^{\text{exp}}(\bar{K}\,K\,\pi) << \Gamma^{\text{exp}}(a_{2}(1320)\,\pi)$: this feature is in agreement with the experiment upon identifying $\bar{K}\,K\,\pi\approx\bar{K}^{\ast}(892)\,K+c.c.$.
		
		\item For what concerns $\eta_{2}(1870)$, the PDG reports $\Gamma^{\text{exp}} (a_{2}(1320)\,\pi) /\Gamma^{\text{exp}}(f_{2}(1270)\,\eta) = 1.7\pm0.4,$ i.e. nonzero. This result excludes $\beta_{pt}=0.0^{\circ}$ (for which the theoretical ratio would vanish). Yet, for $\beta_{pt}=14.8^{\circ}$ a large ratio is obtained ($\approx12$). However, while all experiments do agree that this ratio is sizable, there is a clear disagreement on its actual magnitude. In the next two subsections [(\ref{subsec:large_strange_nonstrange_mixing}) and (\ref{subsec:branching})] we describe this issue in detail.
		
		\item For what concerns $K^{\ast}(892)K+c.c.$ of $\eta_{2}(1870)$, there is also a disagreement with the theoretical results of Tab.\ (\ref{tab:ergebnisisoscalar}) and the experiment. Namely, the theoretical predictions are large. This is understandable, because for a small $\beta_{pt}$ the resonance $\eta_{2}(1870)$ is dominated by its $\bar{s}s$ component. However, the kaonic channel has not been seen in the experiments. Also this aspect is analyzed in the upcoming subsection.
	\end{enumerate}
	
	In conclusion, various inconsistencies with experimental data exist: a small mixing angle must be rejected.


\subsection{Large strange-nonstrange mixing angle}
\label{subsec:large_strange_nonstrange_mixing}

\begin{figure}[H] 
		\begin{subfigure}[b]{\textwidth}
			\centering
			\includegraphics[scale=0.65]{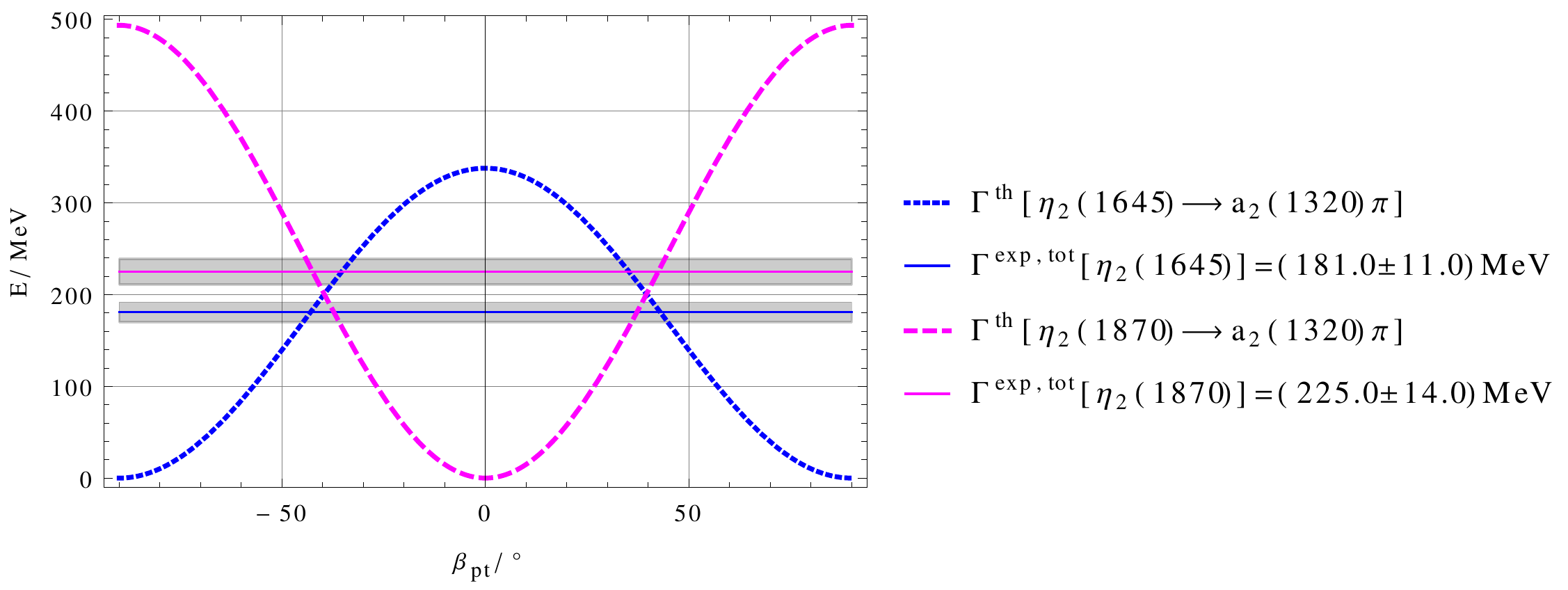}
		\end{subfigure}
		\begin{subfigure}[b]{\textwidth}
			\centering
			\hspace{-5.5mm}
			\includegraphics[scale=0.65]{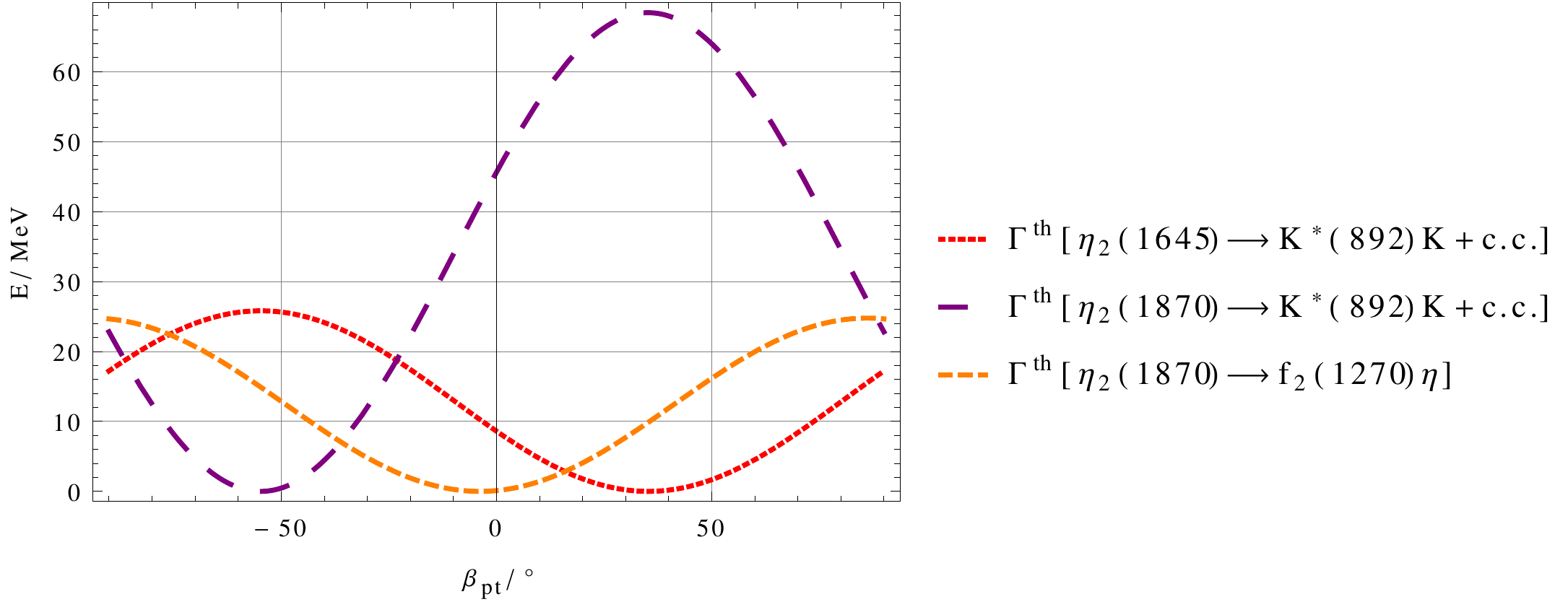}
		\end{subfigure}
		\caption{Upper panel: $\eta_2(1645)\rightarrow a_{2}(1320)\,\pi$ and $\eta_2(1870)\rightarrow a_{2}(1320)\,\pi$ as function of $\beta_{pt}.$ The bands correspond to the total decay widths of $\eta_2(1645)$ and $\eta_2(1870)$. Lower panel: $\eta_2(1645)\rightarrow K^{\ast}(892)\,K,$ $\eta_2(1870)\rightarrow K^{\ast}(892)\,K,$ and $\eta_2(1870)\rightarrow f_{2}(1270)\,\eta$ as function of $\beta_{pt}.$ Only for $\beta_{pt} \approx -40^{\circ}$ the decay $\eta_2(1870)\rightarrow K^{\ast}(892)\,K$ is suppressed. For further details, see discussion in the main text.}
		\label{fig:plots}
	\end{figure}

	Models based on mesonic $\bar{q}q$ nonets and flavour symmetry have proven to be successful, as the clear example of tensor mesons shows \cite{tensor}. Yet, the results of the previous section shows that the case of pseudotensor mesons is more complicated than expected. As an immediate next step, we leave the mixing angle $\beta_{pt}$ unconstrained and test if there is an -- even large -- value which allows for a correct description of all known experimental data.\\
	\\		
	In Fig.\ (\ref{fig:plots}), upper panel, we plot the theoretical decay widths $\eta_2(1645)\rightarrow a_{2}(1320)\,\pi$ and $\eta_2(1870)\rightarrow a_{2}(1320)\,\pi$ as a function of $\beta_{pt}.$ Only for $\beta_{pt}\approx\pm40^{\circ}$ the theoretical decay of $\eta_2(1645)$ and $\eta_2(1870)$ are comparable with the corresponding total experimental widths. Namely, for values of $\beta_{pt} \in [-40^{\circ},\,+40^{\circ}]$ the decay $\eta_2(1645)\rightarrow a_{2}(1320)\,\pi$ exceeds clearly the total decay width of $\eta_2(1645)$ [which is constrained experimentally to be $(181\pm11)\,\,$MeV], while for $\beta_{pt} \in [-90^{\circ},\,-45^{\circ}] \cup [+45^{\circ},\,+90^{\circ}]$ the decay $\eta_2(1645)\rightarrow a_{2}(1320)\,\pi$ exceeds the total decay width of $\eta_2(1870)$ [which is given experimentally by $(225\pm14)\,\,$MeV]. Hence, a large mixing angle $\beta_{pt}\approx\pm40^{\circ}$ is a necessary condition for a consistent description of data.
	
	In Fig.\ (\ref{fig:plots}), lower panel, we show three further decay channels as function of $\beta_{pt}$: $\eta_2(1645)\rightarrow K^{\ast}(892)\,K,$ $\eta_2(1870)\rightarrow K^{\ast}(892)\,K,$ and $\eta_2(1870)\rightarrow f_{2}(1270)\,\eta.$ In particular, the channel $\eta_2(1870)$ $\rightarrow K^{\ast}(892)\,K$ is crucial for our purposes: a negative and large mixing angle, $\beta_{pt}\approx(-40^{\circ},-50^{\circ}),$ is needed to obtain a small partial decay width (this is a consequence of destructive interference). This is necessary to be in agreement with the fact that this decay channel has not been seen in experiments.\\
	Combining the results of both panels, we realize that a large and negative mixing angle is the only possibility. It is instructive to make a definite choice: this is done in Tab.\ (\ref{tab:ergebnisisoscalar-42}) for the illustrative value $\beta_{pt}=-42^{\circ}$. It is visible that all theoretical results are in good agreement with the experimental data.

	\begin{table}[H]
		\centering
		\renewcommand{\arraystretch}{1.1}
		\begin{tabular}[c]{|l|c|c|}
			\hline
			Decay process & Theory (MeV) & Experiment (MeV) \\
			& $(\beta_{pt} = -42^{\circ})$ & \\
			\hline\hline
			$\eta_{2}(1645) \rightarrow\bar{K}^{*}(892)\,K + c.c. $ & $24.7$ & seen \\
			\hline
			$\eta_{2}(1645) \rightarrow a_{2}(1320)\,\pi$ & $186.5$ & \\
			\hline
			$\eta_{2}(1645) \rightarrow\bar{K}_{2}^{*}(1430)\,K + c.c. $ & $0$ & \\
			\hline
			$\eta_{2}(1645) \rightarrow f_{2}(1270)\,\eta$ & $0$ & not seen \\
			\hline
			$\eta_{2}(1645) \rightarrow f_{2}(1270)\,\eta^{\prime}(958) $ & $0$ & \\
			\hline
			$\eta_{2}(1645) \rightarrow f^{\prime}_{2}(1525)\,\eta$ & $0$ & \\
			\hline
			$\eta_{2}(1645) \rightarrow f^{\prime}_{2}(1525)\,\eta^{\prime}(958) $ & $0$ & \\
			\hline\hline
			$\eta_{2}(1870) \rightarrow\bar{K}^{*}(892)\,K + c.c. $ & $3.3$ & \\
			\hline
			$\eta_{2}(1870) \rightarrow a_{2}(1320)\,\pi$ & $221.0$ & \\
			\hline
			$\eta_{2}(1870) \rightarrow\bar{K}_{2}^{*}(1430)\,K + c.c. $ & $0$ & \\
			\hline
			$\eta_{2}(1870) \rightarrow f_{2}(1270)\,\eta$ & $9.4$ & \\
			\hline
			$\eta_{2}(1870) \rightarrow f_{2}(1270)\,\eta^{\prime}(958) $ & $0$ & \\
			\hline
			$\eta_{2}(1870) \rightarrow f^{\prime}_{2}(1525)\,\eta$ & $0$ & \\
			\hline
			$\eta_{2}(1870) \rightarrow f^{\prime}_{2}(1525)\,\eta^{\prime}(958) $ & $0$ & \\
			\hline
		\end{tabular}
		\caption{Decays of $I=0$ pseudotensor states. The total decay widths are $\Gamma_{\eta_{2}(1645)}^{\text{tot}}=(181\pm11)\,\,\text{MeV}$ and $\Gamma_{\eta_{2}(1870)}^{\text{tot}}=(225\pm14)\,\,\text{MeV}$.}
		\label{tab:ergebnisisoscalar-42}
	\end{table}

	Some additional remarks are in order:

	\begin{enumerate}
		\item Our model is based on exact flavour symmetry and only large-$N_{c}$ dominant terms are retained. An agreement at the $5 - 10 \%$ level is expected. In order to achieve a better precision, one should include further terms (see above).
		
		\item The experimental total width of $\eta_2(1645)$ is $(181\pm11)\,\,$MeV, while	the the theoretical result for $\beta_{pt}=-42^{\circ}$ reads $211$ MeV [from Tab.\ (\ref{tab:ergebnisisoscalar-42})]. Taking into account the uncertainty on the coupling constant (keeping $\beta_{pt}=-42^{\circ}$ fixed), one obtains $(211.2\pm13)\,\,$MeV. Obviously, this is only an underestimation of the full error, but it shows that theory and experiment are compatible. The use of remark (1) would improve the agreement.
				
		\item The experimental total width of $\eta_2(1870)$ is $(225\pm14)\,\,$MeV; the corresponding theoretical width for $\beta_{pt}=-42^{\circ}$ reads $233.7\,\,$MeV. Including errors on the coupling constants implies $(233.7\pm15)\,\,$MeV. Also in this case, there is agreement of theory and experiment.
		
		\item The suppressed decay of $\eta_2(1870)$ into $K^{\ast}(892)\,K$ is the result of a destructive interference due to a large and negative strange-nonstrange mixing angle (similar to the one in the pseudoscalar sector). 
	\end{enumerate}

	In conclusion, it is possible to interpret the resonances $[\pi_{2}(1670),\,K_{2}(1770),\,\eta_{2}(1645),\,\eta_{2}(1870)]$ as the ground-state $\bar{q}q$ pseudotensor mesons nonet if a large and negative mixing angle of about $-42^{\circ}$ is considered. In this respect, there is --  at the stage of the present experimental knowledge -- no need to include further additional fields, such as an hybrid pseudotensor state, in the model. (In principle, also a pseudotensor glueball could mix. At the present stage, its mass, as predicted by lattice QCD, is $\approx 3\,\,$GeV \cite{mainlattice}, therefore too large for a sizable effect on decay patterns, see Sec.\ \ref{sec:glueball} for predictions of pseudotensor glueball decays). Yet, there are some experimental conflicting results on branching ratios which we discuss in the next subsection.


\subsection{Branching ratios of $\eta_{2}(1870)$}
\label{subsec:branching}

	We discuss the branching ratios of the resonance $\eta_{2}(1870)$. The experimental information is summarized in Tab.\ (\ref{tab:branching}).
		
	\begin{table}[H] 
		\centering
		\renewcommand{\arraystretch}{1.5}
		\begin{tabular}[c]{|l|c|c|c|}
			\hline
			\multicolumn{4}{|c|}{$\eta_{2}(1870)$} \\
			\hline\hline
			Branching ratio & Theory & Experiment & Collaboration \\
			\hline
			\multirow{4}{*}{$\Gamma^{\text{exp}}(a_2(1320)\,\pi)/\Gamma^{\text{exp}}(f_2(1270)\,\eta)$} &	\multirow{4}{*}{$\approx 23.5$} & $\boldsymbol{1.7 \pm0.4}$ & average by \cite{pdg} \\
			\cline{3-4}
			&  & $1.60 \pm0.4$ & Anisovich et al. \cite{anisovich} \\
			\cline{3-4}
			&  & $20.4 \pm6.6$ & Barberis et al. \cite{barberis} \\
			\cline{3-4}
			&  & $4.1 \pm2.3$ & Adomeit \cite{adomeit} \\
			\hline
			$\Gamma^{\text{exp}}(a_{2}(1320)\,\pi)/\Gamma^{\text{exp}}(a_{0}(980)\,\pi)$ &  & $\boldsymbol{32.6\pm12.6}$ & Barberis et al. \cite{barberis} \\
			\hline
			$\Gamma^{\text{exp}}(a_{0}(980)\,\pi)/\Gamma^{\text{exp}}(f_{2}(1270)\,\eta)$ & & $\boldsymbol{0.48\pm0.45}$ & Barberis et al. \cite{barberis} \\
			\hline
		\end{tabular}
		\caption{Theoretical and experimental branching ratios for $\eta_{2}(1870)$. The mixing angle $\beta_{pt}\approx-42^{\circ}$ is used for theoretical predictions. Bold numbers are also bold in PDG \cite{pdg}.}
		\label{tab:branching}%
	\end{table}
	
	Some of the measurements of the branching ratios $\Gamma^{\text{exp}} (a_{2}(1320)\,\pi)/\Gamma^{\text{exp}} (f_{2}(1270)\,\eta)$ and the average taken by \cite{pdg} are not in agreement. The value of Barberis et al. (for the WA102 collaboration) is much larger ($20.4\pm6.6$) than the others, and was criticized in the recent reanalysis of Anisovich et al. \cite{anisovich}. Actually, the PDG value $1.7\pm0.4$ seems to be solely derived by the result of Anisovich et al. \cite{anisovich} ($1.6\pm0.4$), but it is not clear why the central values differ.\\
	It is also instructive to take the product of the bottom two branching ratios from Tab.\ (\ref{tab:branching}), which should give an estimate for the branching ratio $\Gamma^{\text{exp}} (a_{2}(1320)\,\pi)/\Gamma^{\text{exp}} (f_{2}(1270)\,\eta)$,
	
	\begin{align}
		& \big[ \Gamma(a_{2}(1320)\,\pi)/\Gamma(f_{2}(1270)\,\eta) \big]^{\ast} = \vphantom{\frac{1}{2}}
		\\
		=\, & \big[ \Gamma^{\text{exp}}(a_{2}(1320)\,\pi) / \Gamma^{\text{exp}}(a_{0}(980)\,\pi) \big] \cdot \big[ \Gamma^{\text{exp}}(a_{0}(980)\,\pi)) / \Gamma^{\text{exp}}(f_{2}(1270)\,\eta) \big] \approx \vphantom{\frac{1}{2}} \nonumber
		\\
		\approx\, & \big(16 \pm 16 \big)^\ast \,. \vphantom{\frac{1}{2}} \nonumber
		\\
		&{}^\ast(\text{derived from other exp. branchings}) \vphantom{\frac{1}{2}} \nonumber
	\end{align}
	
\begin{figure}[H] 
		\centering
		\includegraphics[scale=0.75]{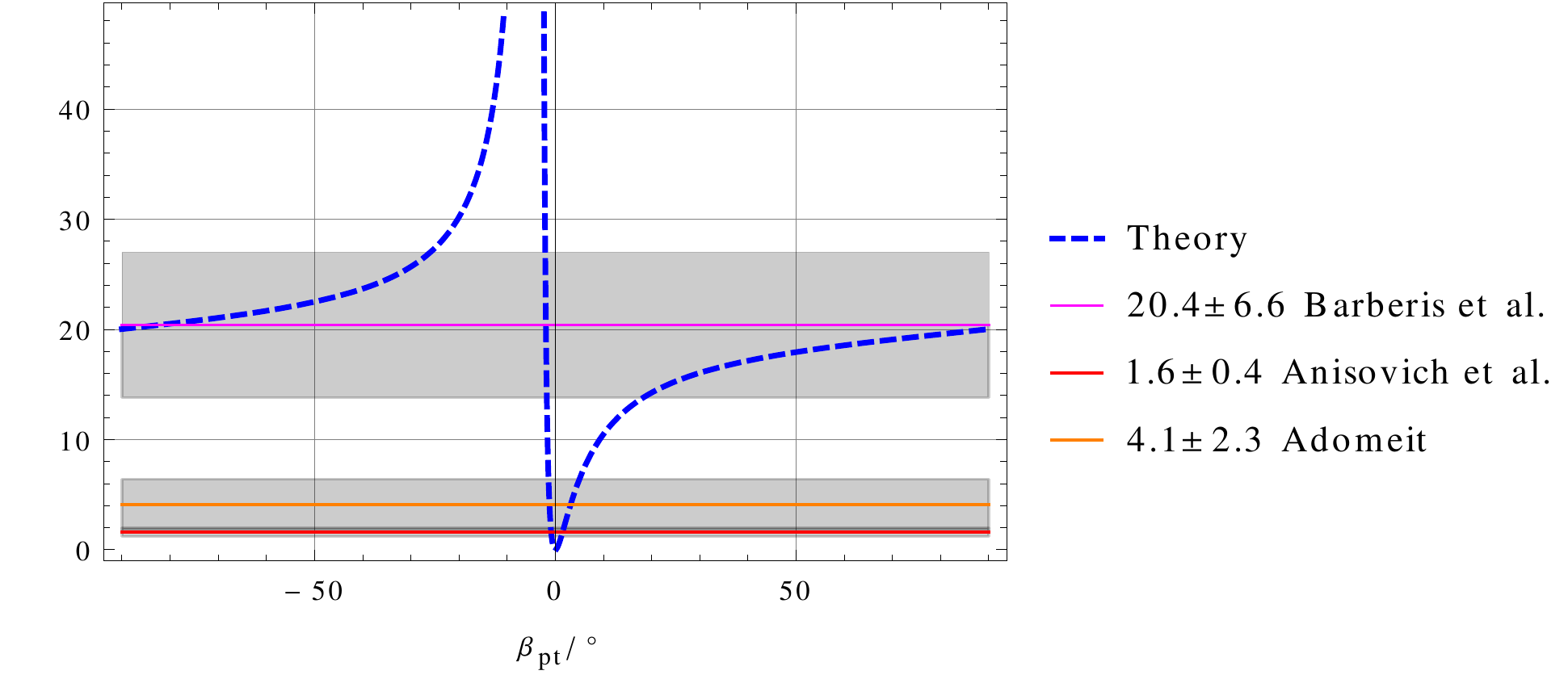}
		\caption{$\beta_{pt}$-dependency of the branching ratio $\Gamma(a_{2}(1320)\,\pi)/\Gamma(f_{2}(1270)\,\eta)$ of $\eta_2(1870)$. The bands correspond to the experimental values, see Tab.\ (\ref{tab:branching}).}
		\label{fig:plot2}
	\end{figure}

	Because of the large errors, this result is in principle compatible with both Anisovich and Barberis, yet the central value is much closer to the latter. New experimental results with a smaller error for the ratio $\Gamma^{\text{exp}}(a_{0}(980)\,\pi ))/\Gamma^{\text{exp}}(f_{2}(1270)\,\eta)$ would be very useful.\\
	\\
	Next, we turn to our prediction (obtained by using $\beta_{pt}=-42^{\circ}$ in our model). We added this result to Tab.\ (\ref{tab:branching}). For $[\eta_{2}(1870)\rightarrow a_{2}(1320)\,\pi]/[\eta_{2}(1870)\rightarrow f_{2}(1270)\,\eta]$ it reads $\approx 23.5$, hence it strongly supports the branching ratio measured by Barberis \cite{barberis}. [The bottom two ratios are not directly included in our model, because $a_0(980)$, as well as all scalar mesons, are not part of it. Hence, the corresponding slots in Tab.\ (\ref{tab:branching}) are empty.]\\
	For better understanding of our prediction ($\approx 23.5$), it is instructive to have a closer look at the theoretical expression for the branching ratio $\Gamma^{th}(a_{2}(1320)\,\pi)/\Gamma^{th}(f_{2}(1270)\,\eta)$ of $\eta_{2}(1870)$. Using Eqs.\ (\ref{eq:txp}) and (\ref{eq:txplagexp}) we find,
	
	\begin{align}
		& \Gamma^{th}(a_{2}(1320)\,\pi)/\Gamma^{th}(f_{2}(1270)\,\eta) = \nonumber
		\\
		=\, &  \frac{3\,\sin^{2}\beta_{pt}}{\big[  -\sin\beta_{pt}\cos\beta_{t} \cos\beta_{p}+\sqrt{2}\cos\beta_{pt}\sin\beta_{t}\sin\beta_{p}\big]^{2}} \left[ \frac{4\,\frac{p_1^{5}}{m_{a_{2}}^{4}}+30\,\frac{p_1^{3}}{m_{a_{2}}^{2}}+45\,p_1}{4\,\frac{k_1^{5}}{m_{f_{2}}^{4}}+30\,\frac{k_1^{3}}{m_{f_{2}}^{2}}+45\,k_1}\right]  \text{ }\,,\label{br}
	\end{align}
	
	where $p_1 = k_{f}(m_{\eta(1870)},m_{a_{2}},m_{\pi})$ and $k_1 = k_{f}(m_{\eta(1870)},m_{f_{2}},m_{\eta}).$ Since $\beta_{t}$ is very small ($\beta_{t}=3.2^{\circ}$), for $\left\vert \beta_{pt}\right\vert $ sufficiently large ($\gtrsim30^{\circ}$) the first term in Eq.\ (\ref{br}) can	be approximated by $3/\cos^{2}\beta_{p}$, which is independent on $\beta_{pt}.$ The complete behavior of ratio is plotted as function of $\beta_{pt}$ in Fig.\ (\ref{fig:plot2}).
	
	One can see that an agreement with Barberis et al. \cite{barberis} is reached for a quite broad range of angles (and -- most importantly -- fits well with $\beta_{pt}\approx -40^{\circ}$). An agreement with Anisovich et al. \cite{anisovich} is only realized for values very close $\beta_{pt} \approx 0^{\circ}$, which is however excluded by the other constraints studied in Subsecs.\ \ref{subsec:small_strange_nonstrange_mixing} and \ref{subsec:large_strange_nonstrange_mixing}.\\
	\\
	In conclusion of the present discussion, $\eta_{2}(1645)$ and $\eta_{2}(1870)$ can be considered as the members of the lowest quark-antiquark nonet only if a large-value ($\approx23$) of the branching ratio $\Gamma(a_{2}(1320)\,\pi/\Gamma(f_{2}(1270)\,\eta))$ of $\eta_{2}(1870)$ holds. Within this respect, the result of Anisovich (ratio $1.6\pm0.4$) would point to a different nature of the resonance $\eta_{2}(1870)$ (see the discussion in Ref. \cite{bing}).\\
	A new experimental study of the resonance $\eta_{2}(1870)$ is necessary to understand which is the correct experimental value. In a near future, both experiments GlueX \cite{gluex1,gluex2,gluex3} and CLAS12 \cite{clas12} (with more emphasis on the former) can study such mesons via photoproduction, hence shading light on this puzzle of mesonic physics. In general, a new determination by GlueX and CLAS12 of the whole pseudotensor sector would be of great help.


\section{Pseudotensor glueball}
\label{sec:glueball}

	Glueballs, bound states formed solely by gluons, have not yet been discovered experimentally (for reviews see Refs.\ \cite{decuplet1,decuplet2,decuplet3,decuplet4,decuplet5,decuplet6,staninew,walaa1,walaa2,Amsler:1995td}). While some experimental candidates for low-lying glueballs have been proposed, the	situation above $3\,\,$GeV is still unsatisfactory. It is then useful to make some (qualitative) predictions of the decay channels of such glueballs in order to help the future experimental identification of candidates.\\
	Among the expected glueballs above 3 GeV, also a glueball with pseudotensor	quantum numbers is expected. This (yet hypothetical) state has a lattice-predicted mass of about $3.04\,\,$GeV \cite{mainlattice,gregory} (for general reviews on glueballs, see Refs. \cite{ochs,vento,crede}). The effective Lagrangian(s) describing the decays of this glueball can be obtained by Eqs.\ (\ref{eq:lagtvp}) and (\ref{eq:lagtxp}) by recalling that each glueball is flavour-blind (see, for instance, the analogous case of the tensor glueball studied in\cite{tensor}). The coupling of the glueball to vector-pseudoscalar pairs is expected to be proportional to $G_{\mu\nu} \mathrm{Tr}\big\{ \big[ V^{\mu} ,\, (\partial^{\nu}P) \big]_{-} \big\}=0,$ hence vanishes. We then expect that

	\begin{align}
		\Gamma_{G\rightarrow VP}=0
	\end{align}

	for all vector-pseudoscalar channels. This is by itself an important information in the identification of possible candidates. In particular, we predict that $\Gamma_{G\rightarrow\rho\pi}=\Gamma_{G\rightarrow K^{\ast}(892)K}=0.$\\
	The coupling of the glueball to tensor-pseudoscalar pairs is given by

	\begin{align}
		\mathcal{L}_{GXP}=c_{GXP}\, G_{\mu\nu} \mathrm{Tr} \big( \big\{ X^{\mu\nu} ,\, P \big\} _{+} \big) \neq 0 \,, \label{eq:glueball_lagrangian}
	\end{align}
	
	where $c_{GXP}$ denotes is the corresponding coupling constant (with dimension energy). The latter cannot be fixed at present because no information on the full width of this glueball state is available (according to large-$N_{c}$ it should be smaller than ordinary mesonic states). [The expanded version of the Lagrangian (\ref{eq:glueball_lagrangian}) is shown in App.\ \ref{eq:gxplagexp}.]\\
	We produce an estimate for ratios of decays, which can be easily calculated following the same steps of the previous sections. They are reported in Tab.\ (\ref{tab:branching_glueball}). The glueball mass is assumed to be $3040\,\,$MeV sharp.
	
	\begin{table}[H] 
		\centering
		\renewcommand{\arraystretch}{1.5}
		\begin{tabular}[c]{|l|c|}
			\hline
			\multicolumn{2}{|c|}{``$G_2(3040)$''} \\
			\hline\hline
			Branching ratio & Theory \\
			\hline
			$\Gamma^{\text{th}}(a_{2}(1320)\,\pi)/\Gamma^{\text{th}}(K_2^{\ast}(1430)\,K + c.c.)$ & 0.9 \\
			\hline
			$\Gamma^{\text{th}}(a_{2}(1320)\,\pi)/\Gamma^{\text{th}}(f_2(1270)\,\eta)$ & 6.0 \\
			\hline
			$\Gamma^{\text{th}}(a_{2}(1320)\,\pi)/\Gamma^{\text{th}}(f_2(1270)\,\eta^\prime(958))$ & 8.5 \\
			\hline
			$\Gamma^{\text{th}}(a_{2}(1320)\,\pi)/\Gamma^{\text{th}}(f^\prime_2(1525)\,\eta)$ & 9.0 \\
			\hline
			$\Gamma^{\text{th}}(a_{2}(1320)\,\pi)/\Gamma^{\text{th}}(f^\prime_2(1525)\,\eta^\prime(958))$ & 11.0 \\
			\hline
		\end{tabular}
		\caption{Theoretical branching ratios for the pseudotensor glueball ``$G_2(3040)$''.}
		\label{tab:branching_glueball}
	\end{table}
	
	It is visible that the decays into $K_{2}^{\ast}(1430)\,K$ and $a_{2}(1320)\,\pi$ are the largest (they are enhanced by isospin factors). It must be stressed that the results of Tab.\ (\ref{tab:branching_glueball}) are determined from the single large-$N_{c}$ and flavour-invariant dominant term of Eq.\ (\ref{eq:glueball_lagrangian}) (the constant $c_{GXP}$ scales as $N_{c}^{-1}$). In line with the expansions described in Sec.\ \ref{sec:themodel}, further terms proportional $c_{GXP}\,\delta_{s}$ as well as $\tilde{c}_{GXP} \propto N_{c}^{-2}$ and so on, are expected. Similarly, a small but non-vanishing decay's amplitude of the pseudotensor glueball into a vector-pseudoscalar pair proportional to $\delta_{s}$ also emerges. All these subleading effects will be important when a suitable glueball's candidate will be experimentally observed.\\
	The search of glueballs between $2.5 - 3\,\,$GeV is currently ongoing at BESIII experiment \cite{bes1,bes2} and will be one of the main subjects of the planned PANDA experiment at the FAIR facility in Darmstadt \cite{panda}. In particular, at PANDA the pseudotensor glueball can be directly formed by proton-antiproton fusion, hence it can be investigated in detail. The here discussed theoretical branching ratios provide help toward the identification of possible candidates by looking at their decay patterns. The full decay width cannot be obtained, however according to large-$N_{c}$ \cite{largen1,largen2,largen3} it should be smaller than conventional quark-antiquark states. A value of about few MeV ($\approx10$ MeV) seems to be a rational guess, see the discussion in Ref. \cite{julia}.


\section{Discussions and conclusions}
\label{sec:conclusions}

	In this work we have studied the phenomenology of the ground-state $\bar{q}q$ pseudotensor meson nonet, identified with the resonances $[\pi_{2}(1670),K_{2}(1770),\eta_{2}(1645),\eta_{2}(1870)]$, by using a quantum field theoretical approach. Two effective interaction Lagrangians which couple pseudotensor states to pseudoscalar, vector and tensor ones were constructed and used to study decays and mixing.\\
	The resonances $\pi_{2}(1670)$ and $K_{2}(1770)$ fit very well into the $\bar{q}q$ picture. However, the isoscalar states $\eta_{2}(1645),\eta_{2}(1870)$ are  more subtle. A small mixing angle between purely nonstrange and strange state is at odd with the experimental data [$\eta_{2}(1645) \rightarrow a_{2}(1320)\,\pi$ is larger than $300\,\,$MeV and overshoots the total experimental width].  A detailed study of the isoscalar sector shows that a good agreement with data is possible if the strange-nonstrange mixing angle is large and negative [approximately $-42^{\circ},$ see Fig.\ (\ref{fig:plots})].\\
	There is however an issue that still needs to be clarified: the ratio $[ \eta_{2}(1870) \rightarrow a_{2}(1320)\, \pi ]/$ $[ \eta_{2}(1870) \rightarrow f_{2}(1270)\, \eta ]$ reads $20.4\pm6.6$ by Barberis \cite{barberis} and $1.6\pm0.4$ by Anisovich \cite{anisovich}. PDG \cite{pdg} lately opted for the latter result ($1.7 \pm 0.4$, with a slightly modified central value). Our study shows that a quarkonium interpretation of $\eta_{2}(1870)$ implies a ratio of about $\approx 23$, hence in good agreement with Barberis but in disagreement with Anisovich. In this respect, a future experimental clarification of this issue is compelling. There are mainly two possible outcomes:
	
	\begin{itemize}
		\item[a)] The ratio quoted by Barberis turns out to be correct, then we have good candidates for a ground-state pseudoscalar meson nonet. However, the large mixing angle (comparable to the one in the pseudoscalar sector) would be a mystery which will deserve a more profound study. Namely, besides the pseudoscalar sector which is affected by the chiral anomaly, all strange-nonstrange mixing angles are small (vector, axial-vector, tensor sectors \cite{pdg,klemptrev,tensor,ciricgliano}). Where would such a large mixing in the pseudotensor sector come from? Would the two-gluon exchange diagram be also enhanced in the pseudotensor channel (a side-effect of the chiral anomaly)?
		
		\item[b)] If, on the contrary, the result of Anisovich shall be confirmed, an understanding of the low-lying pseudotensor states as a standard quark-antiquark nonet would be hard. Modifications would be needed. For instance, one could include additional states in the current approach (such as an hybrid pseudotensor state) which can mix with ordinary $\bar{q}q$ states and change their decay ratios. In such a scenario one would need to consider a decuplet of states (9 standard quarkonia and one additional hybrid state, similar to the decuplet study of scalar states between $1.3 - 1.8\,\,$GeV \cite{decuplet1,decuplet2,decuplet3,decuplet4,decuplet5,decuplet6,Amsler:1995td}). We leave such an analysis for the future, if there will be compelling evidence that the easiest scenario with a low-lying $\bar{q}q$ nonet fails (this is not yet the case, as we have discussed in this work).
	\end{itemize}
	
	As a byproduct of our study, we have also studied the decays of a putative pseudotensor glueball state with a mass of $3.04\,\,$GeV. We find that at leading order it does not decay into pseudoscalar-vector meson pairs (such as $\rho(770)\,\pi$ and $K^{\ast}(892)\,K + c.c.$). On the contrary, sizable decays into $K^{\ast}_{2}(1430)\,K + c.c.$ and $a_{2}(1320)\,\pi$ are expected.\\
	Further improvements of our model (besides the enlargement to a decuplet of	states mentioned above) is the inclusion of large-$N_{c}$ suppressed terms and terms breaking flavor symmetry.  Moreover, tensor and pseudotensor mesons can also build a chiral multiplet. They could be then coupled to the extended Linear Sigma Model in order to test chiral symmetry (and its spontaneous	breaking) in the (pseudo)tensor sector. Also the study of (pseudo)tensor glueballs can be extended in this way. As shown in Ref. \cite{staninew,julia,walaa1,walaa2}, chiral symmetry imposes further constraints on the decays of glueballs, hence improving the predictive powers of hadronic models. From the experimental side, new results and determinations of the pseudotensor mesons $[ \pi_{2}(1670) ,\, K_{2}(1770) ,\, \eta_{2}(1645) ,\, \eta_{2}(1870) ]$ are needed to test the standard basic quark-anti-quark scenario and the existence or not of an enhanced mixing in the isoscalar sector. The upcoming experiments GlueX and CLAS12 experiments at Jlab are expected to	provide results in this direction. Indeed, the properites of pseudotensor mesons are also investigated in the	light-meson program of the COMPASS experiment \cite{Krinner:2016kna}.

	\bigskip

	\textbf{Acknowledgments}: The authors thank Dirk~H.~Rischke for useful discussions. F.G. acknowledges support from the Polish National Science Centre (NCN) through the OPUS project nr. 2015/17/B/ST2/01625. A.K. thanks Dennis~D.~Dietrich for instructive discussions.\\
	\\
	The final publication is available at Springer via http://dx.doi.org//10.1140/epja/i2016-16356-x.

\appendix


\section{ Extended form of the Lagrangians of the model}
\label{app:lagrangians}

The explicit expression for the interaction Lagrangian (\ref{eq:lagtvp}) coupling pseudotensor mesons to vector and pseudoscalar ones is given by:

	\begin{align}
		\mathcal{L}_{TVP} & = c_{TVP}\, \mathrm{Tr} \big\{  T_{\mu\nu} \big[  V^{\mu} ,\, (\partial^{\nu}P) \big]_{-} \big\} = \vphantom{\frac{1}{\sqrt{2}}} \label{eq:tvplagexp}
		\\
		& = c_{TVP}\, \Big(  \frac{1}{\sqrt{2}} \pi_{2,\mu\nu}^{0} \Big[ \bar{K}^{\ast0\mu} (\partial^{\nu}K^{0}) - K^{\ast0\mu} (\partial^{\nu}\bar{K}^{0}) + K^{\ast+\mu} (\partial^{\nu}K^{-}) - K^{\ast-\mu} (\partial^{\nu}K^{+}) + \vphantom{\frac{1}{\sqrt{2}}} \nonumber
		\\
		& \qquad\qquad\qquad\qquad\,\, + 2\, \rho^{+\mu} (\partial^{\nu}\pi^{-}) - 2\, \rho^{-\mu} (\partial^{\nu}\pi^{+}) \Big] + \vphantom{\frac{1}{\sqrt{2}}} \nonumber
		\\
		& \qquad\qquad + \pi_{2,\mu\nu}^{+} \Big[ K^{\ast0\mu} (\partial^{\nu} K^{-}) - K^{\ast-\mu} (\partial^{\nu}K^{0}) + \sqrt{2} \rho^{-\mu} (\partial^{\nu}\pi^{0}) - \sqrt{2} \rho^{0\mu} (\partial^{\nu}\pi^{-}) \Big] \vphantom{\frac{1}{\sqrt{2}}} + \nonumber
		\\
		& \qquad\qquad + \pi_{2,\mu\nu}^{-} \Big[ K^{\ast+\mu} (\partial^{\nu}\bar{K}^{0}) - \bar{K}^{\ast0\mu} (\partial^{\nu}K^{+}) + \sqrt{2} \rho^{0\mu} (\partial^{\nu}\pi^{+}) - \sqrt{2}\rho^{+\mu} (\partial^{\nu} \pi^{0}) \Big] + \vphantom{\frac{1}{\sqrt{2}}} \nonumber
		\\
		& \qquad\qquad + K_{2,\mu\nu}^{0} \Big\{ - \frac{1}{\sqrt{2}} \bar{K}^{\ast0\mu} (\partial^{\nu}\pi^{0}) + K^{\ast-\mu} (\partial^{\nu}\pi^{+}) + \frac{1}{\sqrt{2}} \rho^{0\mu} (\partial^{\nu}\bar{K}^{0}) - \rho^{+\mu} (\partial^{\nu}K^{-}) + \vphantom{\frac{1}{\sqrt{2}}} \nonumber
		\\
		& \qquad\qquad\qquad\qquad\, + \frac{1}{\sqrt{2}} \bar{K}^{\ast0\mu} \big[ (\partial^{\nu}\eta) (\cos\beta_{p} - \sqrt{2} \sin\beta_{p} ) + (\partial^{\nu
		}\eta^{\prime}) ( - \sin\beta_{p} - \sqrt{2} \cos\beta_{p} ) \big] + \vphantom{\frac{1}{\sqrt{2}}} \nonumber
		\\
		& \qquad\qquad\qquad\qquad\, + \frac{1}{\sqrt{2}} \big[ \omega^{\mu} ( - \cos\beta_{v} + \sqrt{2} \sin\beta_{v}) + \phi^{\mu} ( \sin\beta_{v} + \sqrt{2} \cos\beta_{v}) \big] (\partial^{\nu}\bar{K}%
		^{0}) \Big\} + \vphantom{\frac{1}{\sqrt{2}}} \nonumber
		\\
		& \qquad\qquad + K_{2,\mu\nu}^{+} \Big\{ \frac{1}{\sqrt{2}} K^{\ast-\mu} (\partial^{\nu}\pi^{0}) + \bar{K}^{\ast0\mu} (\partial^{\nu}\pi^{-}) - \frac{1}{\sqrt{2}} \rho^{0\mu} (\partial^{\nu}K^{-}) - \rho^{-\mu} (\partial^{\nu}\bar{K}^{0}) + \vphantom{\frac{1}{\sqrt{2}}} \nonumber
		\\
		& \qquad\qquad\qquad\qquad\, + \frac{1}{\sqrt{2}} K^{\ast-\mu} \big[ (\partial^{\nu}\eta) ( \cos\beta_{p} - \sqrt{2} \sin\beta_{p} ) + (\partial^{\nu
		}\eta^{\prime}) ( -\sin\beta_{p} - \sqrt{2} \cos\beta_{p} ) \big] + \vphantom{\frac{1}{\sqrt{2}}} \nonumber
		\\
		& \qquad\qquad\qquad\qquad\, + \frac{1}{\sqrt{2}} \big[ \omega^{\mu} ( - \cos\beta_{v} + \sqrt{2} \sin\beta_{v} ) + \phi^{\mu} ( \sin\beta_{v} + \sqrt{2} \cos\beta_{v} ) \big] (\partial^{\nu}K^{-}) \Big\} + \vphantom{\frac{1}{\sqrt{2}}} \nonumber
		\\
		& \qquad\qquad + K_{2,\mu\nu}^{-} \Big\{ - \frac{1}{\sqrt{2}} K^{\ast+\mu} (\partial^{\nu}\pi^{0}) - K^{\ast0\mu} (\partial^{\nu}\pi^{+}) + \frac{1}{\sqrt{2}} \rho^{0\mu} (\partial^{\nu}K^{+}) + \rho^{+\mu} (\partial^{\nu}K^{0}) + \vphantom{\frac{1}{\sqrt{2}}} \nonumber
		\\
		& \qquad\qquad\qquad\qquad\, + \frac{1}{\sqrt{2}} K^{\ast+\mu} \big[ (\partial^{\nu}\eta) ( - \cos\beta_{p} + \sqrt{2} \sin\beta_{p} ) + (\partial^{\nu}\eta^{\prime}) ( \sin\beta_{p} + \sqrt{2} \cos\beta_{p} ) \big] + \vphantom{\frac{1}{\sqrt{2}}} \nonumber
		\\
		& \qquad\qquad\qquad\qquad\, + \frac{1}{\sqrt{2}} \big[ \omega^{\mu} ( \cos\beta_{v} - \sqrt{2} \sin\beta_{v} ) + \phi^{\mu} ( - \sin\beta_{v} - \sqrt{2} \cos\beta_{v} ) \big] (\partial^{\nu}K^{+}) \Big\} + \vphantom{\frac{1}{\sqrt{2}}} \nonumber
		\\
		& \qquad\qquad + \bar{K}_{2,\mu\nu}^{0} \Big\{ \frac{1}{\sqrt{2}} K^{\ast0\mu} (\partial^{\nu}\pi^{0}) - K^{\ast+\mu} (\partial^{\nu}\pi^{-}) - \frac{1}{\sqrt{2}} \rho^{0\mu} (\partial^{\nu}K^{0}) + \rho^{-\mu} (\partial^{\nu}K^{+}) + \vphantom{\frac{1}{\sqrt{2}}} \nonumber
		\\
		& \qquad\qquad\qquad\qquad\, + \frac{1}{\sqrt{2}} K^{\ast0\mu} \big[ (\partial^{\nu}\eta) ( - \cos\beta_{p} + \sqrt{2} \sin\beta_{p} ) + (\partial^{\nu}\eta^{\prime}) ( \sin\beta_{p} + \sqrt{2} \cos\beta_{p} ) \big] + \vphantom{\frac{1}{\sqrt{2}}} \nonumber
		\\
		& \qquad\qquad\qquad\qquad\, + \frac{1}{\sqrt{2}} \big[ \omega^{\mu} ( \cos\beta_{v} - \sqrt{2} \sin\beta_{v} ) + \phi^{\mu} ( - \sin\beta_{v} - \sqrt{2} \cos\beta_{v} ) \big] (\partial^{\nu}K^{0}) \Big\} + \vphantom{\frac{1}{\sqrt{2}}} \nonumber
		\\
		& \qquad\qquad + \frac{1}{\sqrt{2}} \eta_{2,\mu\nu} ( - \cos\beta_{pt} + \sqrt{2} \sin\beta_{pt}) \cdot \vphantom{\frac{1}{\sqrt{2}}} \nonumber
		\\
		& \qquad\qquad\qquad\qquad\,\,\,\,\cdot \big[ \bar{K}^{\ast0\mu} (\partial^{\nu}K^{0}) - K^{\ast0\mu} (\partial^{\nu}\bar{K}^{0}) - K^{\ast+\mu} (\partial^{\nu}K^{-}) + K^{\ast-\mu} (\partial^{\nu}K^{+}) \big] + \vphantom{\frac{1}{\sqrt{2}}} \nonumber
		\\
		& \qquad\qquad + \frac{1}{\sqrt{2}} \eta_{2,\mu\nu}^{\prime} ( \sin\beta_{pt} + \sqrt{2} \cos\beta_{pt}) \cdot \vphantom{\frac{1}{\sqrt{2}}} \nonumber
		\\
		& \qquad\qquad\qquad\qquad\,\,\,\, \cdot \big[ \bar{K}^{\ast0\mu} (\partial^{\nu}K^{0}) - K^{\ast0\mu} (\partial^{\nu}\bar{K}^{0}) - K^{\ast+\mu} (\partial^{\nu}K^{-}) + K^{\ast-\mu} (\partial^{\nu}K^{+}) \big]  \Big) \,. \vphantom{\frac{1}{\sqrt{2}}} \nonumber
	\end{align}
	
	Similarly, the explicit expression of Eq.\ (\ref{eq:lagtxp}) reads:

	\begin{align}
		\mathcal{L}_{TXP} & = c_{TXP}\, \mathrm{Tr} \big( T_{\mu\nu} \big\{ X^{\mu\nu} ,\, P \big\}_{+} \big) = \vphantom{\frac{1}{\sqrt{2}}} \label{eq:txplagexp}
		\\
		& = c_{TXP}\, \Big( \frac{1}{\sqrt{2}} \pi^0_{2,\mu\nu} \big( - \bar{K}^{*0\mu\nu}_2 K^0 - K^{*0\mu\nu}_2 \bar{K}^0 + K^{*+\mu\nu}_2 K^{-} + K^{*-\mu\nu}_2 K^{+} + \vphantom{\frac{1}{\sqrt{2}}} \nonumber
		\\
		& \qquad \qquad \qquad \qquad + 2\, f^{\mu\nu}_2 \cos\beta_t\, \pi^0 - 2\, f^{\prime\mu\nu}_2 \sin\beta_t\, \pi^0 + 2\, a^{0\mu\nu}_2 \eta \cos\beta_p - 2\, a^{0\mu\nu}_2 \eta^\prime \sin\beta_p \vphantom{\frac{1}{\sqrt{2}}} \big) + \nonumber
		\\
		& \qquad \qquad + \pi^{+}_{2,\mu\nu} \big( K^{*0\mu\nu}_2 K^{-} + K^{*-\mu\nu}_2 K^0 + \sqrt{2} f^{\mu\nu}_2 \cos\beta_t \pi^{-} - \sqrt{2} f^{\prime\mu\nu}_2 \sin\beta_t \pi^{-} + \vphantom{\frac{1}{\sqrt{2}}} \nonumber
		\\
		& \qquad \qquad \qquad \qquad  + \sqrt{2} a^{-\mu\nu}_2 \eta \cos\beta_p - \sqrt{2} a^{-\mu\nu}_2 \eta^\prime \sin\beta_p \big) + \vphantom{\frac{1}{\sqrt{2}}} \nonumber
		\\
		& \qquad \qquad + \pi^{-}_{2,\mu\nu} \big( \bar{K}^{*0\mu\nu}_2 K^{+} + K^{*+\mu\nu}_2 \bar{K}^0 + \sqrt{2} f^{\mu\nu}_2 \cos\beta_t\, \pi^{+} - \sqrt{2} f^{\prime\mu\nu}_2 \sin\beta_t\, \pi^{+} + \vphantom{\frac{1}{\sqrt{2}}} \nonumber
		\\
		& \qquad \qquad \qquad \qquad + \sqrt{2} a^{+\mu\nu}_2 \eta \cos\beta_p - \sqrt{2} a^{+\mu\nu}_2 \eta^\prime \sin\beta_p \big) + \vphantom{\frac{1}{\sqrt{2}}} \nonumber
		\\
		& \qquad \qquad + K^0_{2,\mu\nu} \Big\{ - \frac{1}{\sqrt{2}} \bar{K}^{*0\mu\nu}_2 \pi^0 + K^{*-\mu\nu}_2 \pi^{+} - \frac{1}{\sqrt{2}} a^{0\mu\nu}_2 \bar{K}^0 + a^{+\mu\nu}_2 K^{-} + \vphantom{\frac{1}{\sqrt{2}}} \nonumber
		\\
		& \qquad \qquad \qquad \qquad + \frac{1}{\sqrt{2}} \bar{K}^{*0\mu\nu}_2 \big[ \eta ( \cos\beta_p + \sqrt{2} \sin\beta_p ) + \eta^\prime ( -\sin\beta_p + \sqrt{2} \cos\beta_p ) \big] + \vphantom{\frac{1}{\sqrt{2}}} \nonumber
		\\
		&  \qquad \qquad \qquad \qquad + \frac{1}{\sqrt{2}} \big[ f^{\mu\nu}_2 ( \cos\beta_t + \sqrt{2} \sin\beta_t ) + f^{\prime\mu\nu}_2 ( - \sin\beta_t + \sqrt{2} \cos\beta_t ) \big] \bar{K}^{0} \Big\} + \vphantom{\frac{1}{\sqrt{2}}} \nonumber
		\\
		& \qquad \qquad + K^{+}_{2,\mu\nu} \Big\{ \frac{1}{\sqrt{2}} K^{*-\mu\nu}_2 \pi^0 + \bar{K}^{*0\mu\nu}_2 \pi^{-} + \frac{1}{\sqrt{2}} a^{0\mu\nu}_2 K^{-} + a^{-\mu\nu}_2 \bar{K}^{0} + \vphantom{\frac{1}{\sqrt{2}}} \nonumber
		\\
		& \qquad \qquad \qquad \qquad + \frac{1}{\sqrt{2}} K^{*-\mu\nu}_2 \big[ \eta ( \cos\beta_p + \sqrt{2} \sin\beta_p ) + \eta^\prime ( -\sin\beta_p + \sqrt{2} \cos\beta_p ) \big] + \vphantom{\frac{1}{\sqrt{2}}} \nonumber
		\\
		&  \qquad \qquad \qquad \qquad + \frac{1}{\sqrt{2}} \big[ f^{\mu\nu}_2 ( \cos\beta_t + \sqrt{2} \sin\beta_t ) + f^{\prime\mu\nu}_2 ( - \sin\beta_t + \sqrt{2} \cos\beta_t ) \big] K^{-} \Big\} + \vphantom{\frac{1}{\sqrt{2}}} \nonumber
		\\
		& \qquad \qquad + K^{-}_{2,\mu\nu} \Big\{ \frac{1}{\sqrt{2}} K^{*+\mu\nu}_2 \pi^0 + K^{*0\mu\nu}_2 \pi^{+} + \frac{1}{\sqrt{2}} a^{0\mu\nu}_2 K^{+} + a^{+\mu\nu}_2 K^{0} + \vphantom{\frac{1}{\sqrt{2}}} \nonumber
		\\
		& \qquad \qquad \qquad \qquad + \frac{1}{\sqrt{2}} K^{*+\mu\nu}_2 \big[ \eta ( \cos\beta_p + \sqrt{2} \sin\beta_p ) + \eta^\prime ( -\sin\beta_p + \sqrt{2} \cos\beta_p ) \big] + \vphantom{\frac{1}{\sqrt{2}}} \nonumber
		\\
		&  \qquad \qquad \qquad \qquad + \frac{1}{\sqrt{2}} \big[ f^{\mu\nu}_2 ( \cos\beta_t + \sqrt{2} \sin\beta_t ) + f^{\prime\mu\nu}_2 ( - \sin\beta_t + \sqrt{2} \cos\beta_t ) \big] K^{+} \Big\} + \vphantom{\frac{1}{\sqrt{2}}} \nonumber
		\\
		& \qquad \qquad + \bar{K}^{0}_{2,\mu\nu} \Big\{ - \frac{1}{\sqrt{2}} K^{*0\mu\nu}_2 \pi^0 + K^{*+\mu\nu}_2 \pi^{-} - \frac{1}{\sqrt{2}} a^{0\mu\nu}_2 K^{0} + a^{-\mu\nu}_2 K^{+} + \vphantom{\frac{1}{\sqrt{2}}} \nonumber
		\\
		& \qquad \qquad \qquad \qquad + \frac{1}{\sqrt{2}} K^{*0\mu\nu}_2 \big[ \eta ( \cos\beta_p + \sqrt{2} \sin\beta_p ) + \eta^\prime ( -\sin\beta_p + \sqrt{2} \cos\beta_p ) \big] + \vphantom{\frac{1}{\sqrt{2}}} \nonumber
		\\
		&  \qquad \qquad \qquad \qquad + \frac{1}{\sqrt{2}} \big[ f^{\mu\nu}_2 ( \cos\beta_t + \sqrt{2} \sin\beta_t ) + f^{\prime\mu\nu}_2 ( - \sin\beta_t + \sqrt{2} \cos\beta_t ) \big] K^{0} \Big\} + \vphantom{\frac{1}{\sqrt{2}}} \nonumber
		\\
		& \qquad \qquad + \frac{1}{\sqrt{2}} \eta_{2,\mu\nu} \big[ ( \cos\beta_{pt} + \sqrt{2} \sin\beta_{pt} ) \cdot \vphantom{\frac{1}{\sqrt{2}}} \nonumber
		\\
		& \qquad \qquad \qquad \qquad \quad \,\,\, \cdot ( \bar{K}^{*0\mu\nu}_2 K^0 + K^{*0\mu\nu}_2 \bar{K}^0 + K^{*+\mu\nu}_2 K^{-} + K^{*-\mu\nu}_2 K^{+} ) + \vphantom{\frac{1}{\sqrt{2}}} \nonumber
		\\
		& \qquad \qquad \qquad \qquad \quad \,\,\, + 2\, \cos\beta_{pt}\, ( a^{0\mu\nu}_2 \pi^0 + a^{+\mu\nu}_2 \pi^{-} + a^{-\mu\nu}_2 \pi^{+} ) + \vphantom{\frac{1}{\sqrt{2}}} \nonumber
		\\
		& \qquad \qquad \qquad \qquad \quad \,\,\, + 2\, \cos\beta_{pt}\, ( f^{\mu\nu}_2 \cos\beta_t - f^{\prime\mu\nu}_2 \sin\beta_t ) ( \eta \cos\beta_p - \eta^\prime \sin\beta_p ) + \vphantom{\frac{1}{\sqrt{2}}} \nonumber
		\\
		& \qquad \qquad \qquad \qquad \quad \,\,\, + 2 \sqrt{2} \sin\beta_{pt}\, ( f^{\mu\nu}_2 \sin\beta_t + f^{\prime\mu\nu}_2 \cos\beta_t ) ( \eta \sin\beta_p + \eta^\prime \cos\beta_p ) \big] + \vphantom{\frac{1}{\sqrt{2}}} \nonumber
		\\
		& \qquad \qquad + \frac{1}{\sqrt{2}} \eta^\prime_{2,\mu\nu} \big[ ( - \sin\beta_{pt} + \sqrt{2} \cos\beta_{pt} ) \cdot \vphantom{\frac{1}{\sqrt{2}}} \nonumber
		\\
		& \qquad \qquad \qquad \qquad \quad \,\,\, \cdot ( \bar{K}^{*0\mu\nu}_2 K^0 + K^{*0\mu\nu}_2 \bar{K}^0 + K^{*+\mu\nu}_2 K^{-} + K^{*-\mu\nu}_2 K^{+} ) + \vphantom{\frac{1}{\sqrt{2}}} \nonumber
		\\
		& \qquad \qquad \qquad \qquad \quad \,\,\, - 2\, \sin\beta_{pt}\, ( a^{0\mu\nu}_2 \pi^0 + a^{+\mu\nu}_2 \pi^{-} + a^{-\mu\nu}_2 \pi^{+} ) + \vphantom{\frac{1}{\sqrt{2}}} \nonumber
		\\
		& \qquad \qquad \qquad \qquad \quad \,\,\, - 2\, \sin\beta_{pt}\, ( f^{\mu\nu}_2 \cos\beta_t - f^{\prime\mu\nu}_2 \sin\beta_t ) ( \eta \cos\beta_p - \eta^\prime \sin\beta_p ) + \vphantom{\frac{1}{\sqrt{2}}} \nonumber
		\\
		& \qquad \qquad \qquad \qquad \quad \,\,\, + 2 \sqrt{2} \cos\beta_{pt}\, ( f^{\mu\nu}_2 \sin\beta_t + f^{\prime\mu\nu}_2 \cos\beta_t ) ( \eta \sin\beta_p + \eta^\prime \cos\beta_p ) \big] \,.  \vphantom{\frac{1}{\sqrt{2}}} \nonumber
	\end{align}
	
	The explicit form of the glueball interaction Lagrangian (\ref{eq:glueball_lagrangian}) that couples the pseudotensor glueball to tensor and pseudoscalar mesons reads,
	
	\begin{align}
		\mathcal{L}_{GXP} & = c_{GXP}\, G_{\mu\nu} \mathrm{Tr} \big( \big\{ X^{\mu\nu} ,\, P \big\} _{+} \big) = \vphantom{\frac{1}{\sqrt{2}}} \label{eq:gxplagexp}
		\\
		& = 2\, c_{GXP}\, G_{\mu\nu} \big[ a_2^{0\mu\nu} \pi^0 + a_2^{+\mu\nu} \pi^{-} + a_2^{-\mu\nu} \pi^{+} + \vphantom{\frac{1}{\sqrt{2}}} \nonumber
		\\
		& \hphantom{= 2\, c_{GXP}\, G_{\mu\nu} \big[} + \bar{K}_2^{\ast 0 \mu\nu} K^0 + K_2^{\ast 0 \mu\nu} \bar{K}^0 + K_2^{\ast + \mu\nu} K^{-} + K_2^{\ast - \mu\nu} K^{+} + \vphantom{\frac{1}{\sqrt{2}}} \nonumber
		\\
		& \hphantom{= 2\, c_{GXP}\, G_{\mu\nu} \big[} + ( f_2^{\mu\nu} \cos \beta_t - f_2^{\prime\mu\nu} \sin \beta_t ) ( \eta \cos \beta_p - \eta^\prime \sin \beta_p ) + \vphantom{\frac{1}{\sqrt{2}}} \nonumber
		\\
		& \hphantom{= 2\, c_{GXP}\, G_{\mu\nu} \big[} + ( f_2^{\mu\nu} \sin \beta_t + f_2^{\prime\mu\nu} \cos \beta_t ) ( \eta \sin \beta_p + \eta^\prime \cos \beta_p ) \big] \,. \vphantom{\frac{1}{\sqrt{2}}} \nonumber
	\end{align}


\section{Unpolarized invariant decay amplitudes}
\label{app:deacyamplitudes}

	In this appendix we calculate the unpolarized invariant decay amplitudes $\frac{1}{5}\sum|\mathcal{M}|^{2}$ for the decay widths (\ref{eq:tvp}) and (\ref{eq:txp}). We use Feynman rules to translate the Lagrangian densities (\ref{eq:lagtvp}) and (\ref{eq:lagtxp}) into decay amplitudes. Therefore, the trace and matrix structures of the given Lagrangians only modify the specific coupling constants of the single decay channels. Consequently, we work with the simplified Lagrangian densities
	
	\begin{align}
		\mathcal{L}_{TVP} &  =g_{TVP}\, T_{\mu\nu} V^{\mu} (\partial^{\nu}P)\,,\\
		\mathcal{L}_{TXP} &  =g_{TXP}\, T_{\mu\nu} X^{\mu\nu} P\,.
	\end{align}
	
	The Feynman rules are:
	
	\begin{align}
		g & \rightarrow -i\,g \,, \vphantom{T^{\mu\nu} \rightarrow \epsilon^{\mu\nu}(\lambda_t,\,\vec{k}_t)} \nonumber
		\\
		T^{\mu\nu} &  \rightarrow \epsilon^{\mu\nu} (\lambda_{t},\,\vec{k}_{t}) \,, \nonumber
		\\
		X^{\mu\nu} &  \rightarrow \epsilon^{\mu\nu} (\lambda_{x},\,\vec{k}_{x}) \,, \vphantom{T^{\mu\nu} \rightarrow \epsilon^{\mu\nu}(\lambda_t,\,\vec{k}_t)} \nonumber
		\\
		V^{\mu} & \rightarrow \epsilon^{\mu} (\lambda_{v},\,\vec{k}_{v}) \,, \vphantom{T^{\mu\nu} \rightarrow \epsilon^{\mu\nu}(\lambda_t,\,\vec{k}_t)} \nonumber
		\\
		\partial^{\mu} & \rightarrow i\,k_{p}^{\mu} \,, \vphantom{T^{\mu\nu} \rightarrow\epsilon^{\mu\nu}(\lambda_t,\,\vec{k}_t)}
	\end{align}
	
	where $\epsilon^{\mu(\nu)}$ are polarization vectors (tensors), $\lambda$ is the specific polarizations\footnote{For tensors $\lambda =-2,\,-1,\,0,\,+1,\,+2$, whereas for vectors $\lambda=-1,\,0,\,+1$, representing the spin projections, or respective degrees of freedom.}, $k^{\mu}$ is the four-momentum, and $\vec{k}$ is the three-momentum. Hence,
	
	\begin{align}
		i\mathcal{M}_{TVP} & = -i\, g_{TVP}\, \epsilon_{\mu\nu} (\lambda_{t} ,\, \vec{k}_{t})\, \epsilon^{\mu}(\lambda_{v} ,\, \vec{k}_{v}) \, i\, k_{p}^{\nu}	= g_{tvp}\, \epsilon_{\mu\nu}(\lambda_{t} ,\, \vec{k}_{t})\, \epsilon^{\mu} (\lambda_{v} ,\, \vec{k}_{v})\, k_{p}^{\nu}\,,
		\\
		i\mathcal{M}_{TXP} & = -i\, g_{TXP}\, \epsilon_{\mu\nu} (\lambda_{t} ,\, \vec{k}_{t})\, \epsilon^{\mu\nu} (\lambda_{x} ,\, \vec{k}_{x}) \,.
	\end{align}
	
	The unpolarized invariant decay amplitudes is found by taking the square of the above expressions. In addition one has to average over all incoming spin-polarizations and sum up all possible outgoing polarizations. For the decay of pseudotensor resonances into a vector and a pseudoscalar mesons we find:
	
	\begin{align}
		\frac{1}{5} \sum |\mathcal{M}_{TVP}|^{2} & = \frac{g_{TVP}^{2}}{5} \sum_{\lambda_{t}=1}^{5} \sum_{\lambda_{v}=1}^{3} \epsilon_{\mu\nu} (\lambda_{t} ,\, \vec{k}_{t})\, \epsilon_{\alpha\beta} (\lambda_{t} ,\, \vec{k}_{t})\, \epsilon^{\mu} (\lambda_{v} ,\, \vec{k}_{v})\, \epsilon^{\alpha} (\lambda_{v} ,\, \vec{k}_{v})\, k_{p}^{\nu}\, k_{p}^{\beta} = \vphantom{\Big( 2\, \frac{\vec{k}^4}{m_v^2} + 5\,\vec{k}^2 \Big)} \nonumber
		\\
		& = \frac{g_{tvp}^{2}}{15} \Big( 2\,\frac{|\vec{k}|^{4}}{m_{v}^{2}} + 5\,|\vec{k}|^{2} \Big)  \,,
	\end{align}
	
	where the rest frame of the decaying pseudo-tensor meson is considered ($\vec{k}$ denotes the momentum of an outgoing particle: $|\vec{k}|=k_{f}(m_{T},m_{V},m_{P})$). We made use of the following completeness relations
		
	\begin{align}
		\sum_{\lambda=1}^{3} \epsilon_{\mu}(\lambda ,\, \vec{k})\, \epsilon_{\nu}(\lambda ,\, \vec{k}) & = -G_{\mu\nu}\,, \label{eq:completenessvector}
		\\
		\sum_{\lambda=1}^{5} \epsilon_{\mu\nu} (\lambda ,\, \vec{k})\, \epsilon_{\alpha\beta} (\lambda ,\, \vec{k}) & = -\frac{G_{\mu\nu} G_{\alpha\beta}}{3} + \frac{G_{\mu\alpha} G_{\nu\beta} + G_{\mu\beta} G_{\nu\alpha}}{2} \,, \label{eq:completenesstensor}
	\end{align}
	
	where
	
	\begin{align}
		G_{\mu\nu} & = \eta_{\mu\nu} - \frac{k_{\mu} k_{\nu}}{m^{2}} \,.
	\end{align}
	
	(For details, see Ref. \cite{Koenigstein:2015} and refs. therein). In analogy to the previous case, the unpolarized invariant decay amplitude for the decay	of pseudotensor meson into a tensor and a pseudoscalar mesons reads:
	
	\begin{align}
		\frac{1}{5} \sum|\mathcal{M}_{TXP}|^{2} & = \frac{g_{TXP}^{2}}{5} \sum_{\lambda_{t}=1}^{5} \sum_{\lambda_{x}=1}^{5} \epsilon_{\mu\nu} (\lambda_{t} ,\, \vec{k}_{t})\, \epsilon_{\alpha\beta} (\lambda_{t} ,\, \vec{k}_{t})\, \epsilon^{\mu\nu} (\lambda_{x} ,\, \vec{k}_{x})\, \epsilon^{\alpha\beta} (\lambda_{x} ,\, \vec{k}_{x}) = \vphantom{\left( 4\, \frac{|\vec{k}|^4}{m_x^4} + 30\,\frac{|\vec{k}|^2}{m_x^2} +45 \right)} \nonumber
		\\
		& = \frac{g_{TXP}^{2}}{45} \Big(  4\, \frac{|\vec{k}|^{4}}{m_{X}^{4}} + 30\, \frac{|\vec{k}|^{2}}{m_{X}^{2}} + 45 \Big) \,
	\end{align}
	
	where $|\vec{k}|=k_{f}(m_{T},m_{X},m_{P})$.

\end{document}